\documentclass[pra,twocolumn]{revtex4}
\usepackage{amsfonts,amsmath,amssymb,float}
\usepackage{bm,dsfont}
\usepackage{color}
\usepackage{bbm}
\usepackage{longtable}
\usepackage[dvips]{epsfig}
\usepackage{amsmath,amssymb,lscape,float}
\usepackage{hyperref}
\usepackage{listings}

\usepackage{graphicx}
\usepackage{epstopdf}
\usepackage{dcolumn}
\usepackage{bm}
\usepackage{amsfonts,amsmath,amssymb
}

\usepackage{bm,dsfont}
\usepackage{color}
\usepackage{bbm}
\usepackage{longtable}
\usepackage[dvips]{epsfig}
\usepackage{hyperref}
\usepackage{listings}
\usepackage{todonotes}

\newcommand{\bra}[1]{\langle #1|}

\newcommand{\ket}[1]{|#1 \rangle}

\newcommand{\uber}[2]{{{#1}\choose{#2}}}

\begin{document}
\title{Dicke-state preparation through global transverse control of Ising-coupled qubits}
\author{Vladimir M. Stojanovi\'c}
\affiliation{Institut f\"{u}r Angewandte Physik, Technical
University of Darmstadt, D-64289 Darmstadt, Germany}

\author{Julian K. Nauth}
\affiliation{Institut f\"{u}r Angewandte Physik, Technical University of Darmstadt, 
D-64289 Darmstadt, Germany}

\date{\today}
\begin{abstract}
We consider the problem of engineering the two-excitation Dicke state $|D^{3}_{2}\rangle$ in a three-qubit system
with all-to-all Ising-type qubit-qubit interaction, which is also subject to global transverse (Zeeman-type) 
control fields. The theoretical underpinning for our envisioned state-preparation scheme, in which $|000\rangle$ is 
adopted as the initial state of the system, is provided by a Lie-algebraic result that guarantees state-to-state controllability 
of this system for an arbitrary choice of initial- and final states that are invariant with respect to permutations 
of qubits. This scheme is envisaged in the form of a pulse sequence that involves three instantaneous control pulses, 
which are equivalent to global qubit rotations, and two Ising-interaction pulses of finite durations between consecutive 
control pulses. The design of this pulse sequence (whose total duration is $T\approx 0.95\:\hbar/J$, where $J$ is
the Ising-coupling strength) leans heavily on the concept of the symmetric sector, a four-dimensional, permutationally-invariant
subspace of the three-qubit Hilbert space. We demonstrate the feasibility of the proposed state-preparation scheme by 
carrying out a detailed numerical analysis of its robustness to systematic errors, i.e. deviations from the optimal 
values of the eight parameters that characterize the underlying pulse sequence. Finally, we discuss how 
our proposed scheme can be generalized for engineering Dicke states in systems with $N \ge 4$ qubits. For the sake 
of illustration, we describe the preparation of the two-excitation Dicke state $|D^{4}_{2}\rangle$ in a four-qubit 
system.
\end{abstract}
	
\maketitle
\section{Introduction}
Advanced capabilities in quantum-state engineering~\cite{Dogra+:14,Das+:15,Li+Song:15,Kang+:16,Kang+SciRep:16,StojanovicPRL:20,
Peng+:21,StojanovicPRA:21,Zheng++:22,Zhang+:23,Zhang++:23,Song+:17,Erhard+:18,Macri+:18,Zheng+:19,Nogueira+:21,Qiao+:22,
Feng+:22,Pachniak+Malinovskaya:21,Hauck+:21,Hauck+Stojanovic:22,Zheng+:20,Haase+:21,Haase++:22,Nauth+Stojanovic:22,Stojanovic+Nauth:22,
Shao+:23} represent 
one of the crucial prerequisites for the development of next-generation quantum technologies~\cite{Dowling+Milburn:03}. Tantalizing 
achievements have been reported in this thriving research area in recent years, pertaining -- in particular -- to the creation 
of highly entangled quantum states in systems of increasingly large size that belong to various physical platforms for quantum 
computing (QC)~\cite{Haas+:14,Friis+:18,Song+:17}. Two of the most widely discussed classes of such states are maximally-entangled 
multipartite states of $W$~\cite{Duer+:00} and Greenberger-Horne-Zeilinger (GHZ)~\cite{Greenberger+Horne+Zeilinger:89} types. 
Motivated primarily by their proven utility in various quantum-information processing~\cite{NielsenChuangBook} tasks, a multitude 
of proposals have been made in recent years for the efficient generation of both $W$~\cite{Dogra+:14,Das+:15,Li+Song:15,Kang+:16,
Kang+SciRep:16,StojanovicPRL:20,Peng+:21,StojanovicPRA:21,Zheng++:22,Zhang+:23,Zhang++:23} and GHZ~\cite{Song+:17,Erhard+:18,
Macri+:18,Zheng+:19,Nogueira+:21,Qiao+:22,Feng+:22,Pachniak+Malinovskaya:21} states in diverse QC platforms. In addition, 
deterministic interconversions of $W$ and GHZ states have lately also been attracting considerable attention~\cite{Zheng+:20,
Haase+:21,Haase++:22,Nauth+Stojanovic:22,Stojanovic+Nauth:22,Shao+:23}.

Dicke states, originally introduced in connection with the phenomenon of superradiance~\cite{Dicke:54}, represent yet another important 
family of highly entangled multiqubit states. Owing to their favorable properties -- such as robustness to particle loss~\cite{Neven+:18} 
and immunity to collective dephasing noise~\cite{Lidar+Whaley:03} -- they hold promise for emerging quantum-technology applications 
in areas as diverse as quantum networking~\cite{Prevedel+:09}, quantum metrology~\cite{Toth:12}, quantum game theory~\cite{Ozdemir+:07}, 
and quantum combinatorial optimization~\cite{Childs+:02}. The $a$-excitation Dicke state $\ket{D^{N}_{a}}$ of an $N$-qubit system 
is an equal-weight superposition of all $N$-qubit states with Hamming weight of $a$ (i.e. all binary bitstrings of length $N$ with 
exactly $a$ appearances of $1$); this family of states includes $W$ states as its special, single-excitation ($a=1$) case, i.e. 
$\ket{D^{N}_{1}}\equiv\ket{W_{N}}$. Several schemes have heretofore been proposed for the experimental realization of Dicke states 
in various physical platforms, such as trapped ions~\cite{Hume+:09,Ivanov+:13,Lamata+:13}, neutral atoms~\cite{Stockton+:04,Xiao+:07,
Shao+:10}, photons~\cite{Prevedel+:09,Wieczorek+:09}, superconducting qubits~\cite{Wu+:17}, and spin ensembles~\cite{Wang+Terhal:21}.
	
In this paper, we consider the problem of engineering the two-excitation Dicke state $\ket{D^{3}_{2}}$ in a three-qubit system 
with long-ranged (all-to-all) Ising-type ($zz$) interaction between qubits, which are acted upon by global, Zeeman-type
control fields in the transverse ($x$- and $y$) directions. Aside from quantum-technology applications of Dicke states,
the motivation behind the present work stems from the fact that qubit arrays with Ising-type qubit-qubit interaction can be 
realized in several physical platforms for QC~\cite{Solgun+Srinivasan:22}. One familiar example is furnished by nuclear spins, 
i.e. nuclear-magnetic-resonance (NMR) systems~\cite{Vandersypen+Chuang:05,Viola+:01,Zhang+:15}. Another example are 
capacitively-coupled spin qubits of singlet-triplet type~\cite{Shulman+:12}, formed using double quantum dots~\cite{Levy:02}. Finally, 
arrays of optically-trapped neutral atoms interacting via off-resonant dipole-dipole (van-der-Waals) interactions feature 
an all-to-all Ising-type interaction between neutral-atom qubits based on Rydberg states~\cite{Morgado+Whitlock:21,ShiREVIEW:22}.
This last, neutral-atom-based platform offers particularly rich possibilities for quantum-state engineering~\cite{Li+:22}.

A recent result in the realm of Lie-algebraic controllability implies that a qubit array with all-to-all Ising-type interaction 
between qubits, when acted upon by two global transverse control fields, is state-to-state controllable if the initial 
and final state are invariant under an arbitrary permutation of qubits~\cite{Albertini+DAlessandro:18}; moreover, it is important 
to note that both our adopted initial state $|000\rangle$ and the sought-after Dicke state $\ket{D^{3}_{2}}$ are indeed 
permutationally invariant. Thus, this last Lie-algebraic result guarantees the existence of a quantum-control protocol for 
the preparation of the state $\ket{D^{3}_{2}}$ starting from $|000\rangle$ for Ising-coupled qubits with global transverse control. 

Our envisaged state-preparation scheme is based on an NMR-type pulse sequence. This pulse sequence turns out to involve three instantaneous 
($\delta$-shaped) global control pulses, two of which correspond to global qubit rotations around the $y$ axis and the remaining one 
around the $x$ axis, and two Ising-interaction pulses of equal durations between consecutive control pulses. It is worthwhile to mention 
that NMR-type pulse sequences have heretofore been used in various QC problems~\cite{Jones:03,Hill:07,Geller+:10,Ghosh+Geller:10,TanamotoQECC,
Tanamoto+:12,Tanamoto+:13, Stefanatos+Paspalakis:20}, even in the presence of the same (Ising) type of qubit-qubit interaction. For instance, 
they have been utilized for preserving cluster states~\cite{Tanamoto+:12} in measurement-based QC~\cite{Raussendorf+Briegel:01}; the practical 
usefulness of the Ising-type qubit-qubit coupling in this regard stems from the close kinship between its native two-qubit gate and 
controlled-$Z$~\cite{NielsenChuangBook}, the gate used for generating cluster states~\cite{Raussendorf+Briegel:01}.

The design of the envisaged pulse sequence relies heavily on the permutational invariance of the state-preparation problem at hand -- more 
precisely, on the use of the four-dimensional, permutationally-invariant subspace (symmetric sector) of the three-qubit Hilbert space. 
After finding the optimal values of the eight parameters characterizing this pulse sequence (namely, the durations of the interaction pulses, 
the global-rotation angles, which are related to the control-field magnitudes, and, finally, the angles specifying the directions of the 
rotation axes in the $x$-$y$ plane), we carry out a detailed numerical analysis of its sensitivity against systematic errors in the values 
of those parameters. We show that even for fairly large relative errors in different parameter values, one can still retain Dicke-state 
fidelities very close to unity, which speaks in favor of the robustness of the proposed scheme.

The remaining part of this paper is organized in the following manner. In Sec.~\ref{SystemProblem}, the Ising-coupled qubit system 
at hand and the Dicke-state preparation problem to be addressed in what follows are described in detail; the notation to be used 
throughout the paper is also introduced, along with some permutational-symmetry-related considerations. Section~\ref{DickeSequence} 
is set aside for the discussion of the design of the sought-after NMR-type pulse sequence that allows an efficient Dicke-state 
preparation. The obtained results for the idealized pulse sequence, i.e. for the optimal values of its eight characteristic parameters,
are then presented and discussed. Finally, a geometric interpretation of the proposed pulse sequence, based on a dimensional reduction 
of the problem at hand and the concept of the Bloch sphere, is also provided. In Sec.~\ref{PulseSequenceRobust} the robustness of the 
proposed scheme to systematic errors in its characteristic parameters is discussed in great detail. Section~\ref{GeneralizeNgt4}
is concerned with the generalization of the proposed state-preparation scheme to systems with four or more qubits; it starts with 
some general symmetry-related considerations, followed by the demonstration of the pulse sequence for realizing the two-excitation
Dicke state in a four-qubit system. Before closing, in Sec.~\ref{SummConcl} we summarize the main findings of this paper and indicate 
possible directions for future work. The derivation of the time-evolution operators corresponding to the different stages 
of the proposed pulse sequences in the three- and four-qubit cases is relegated to Appendix~\ref{DerivTimeEvol}.

\section{System and Dicke-state preparation problem} \label{SystemProblem}
To set the stage for further discussion, we start by introducing a system of Ising-coupled qubits with global transverse control,
with emphasis on the relevant Lie-algebraic controllability results (Sec.~\ref{SystemHamilt}). We then briefly review some basic 
properties of Dicke states and formulate the state-preparation problem under consideration as a quantum-control problem in a three-qubit 
system (Sec.~\ref{DickeStatePrep}). Finally, we underscore the consequences of the permutational invariance of the system at hand 
for the solution of this quantum-control problem (Sec.~\ref{SymmSector}).

\subsection{Hamiltonian of Ising-coupled qubit arrays with global transverse control} \label{SystemHamilt}
The system at hand is an array of qubits coupled through long-range (all-to-all) Ising ($zz$) coupling,
which are also acted upon by global Zeeman-type control fields in the two transverse ($x$ and $y$) directions. 
Accordingly, the total system Hamiltonian $H(t)=H_{ZZ}+H_C(t)$ is given by the sum of the Ising-interaction
part $H_{ZZ}$ and the control part, i.e.,
\begin{equation} \label{TotalHamiltonian}
H(t)=H_{ZZ} + h_x(t)\mathcal{X}+h_y(t)\mathcal{Y}\:,
\end{equation}
where $h_x(t)$ and $h_y(t)$ are global control fields in the $x$- and $y$ directions, respectively.
The operators $H_{ZZ}$, $\mathcal{X}$, and $\mathcal{Y}$ are given by 
\begin{eqnarray} 
H_{ZZ} &=& J\sum_{1\le n<n'\le N}\:Z_n Z_{n'} \:, \label{LRIsingInt} \\
\mathcal{X} &=& \sum_{n=1}^{N}\:X_n \quad,\quad \mathcal{Y} = 
\sum_{n=1}^{N}\:Y_n\:, \label{ControlOperators}
\end{eqnarray}
where $J$ denotes the Ising coupling strength, while $X_n$, $Y_n$, and $Z_n$ are the Pauli operators 
of the $n$-th qubit ($n=1,\ldots,N$), i.e.
\begin{equation} \label{defineXYZ_n}
\mathbf{X}_n = \mathbbm{1}_2\otimes\ldots\otimes\mathbbm{1}_2\otimes 
\underbrace{\mathbf{X}}_n\otimes\mathbbm{1}_2\otimes\ldots\otimes\mathbbm{1}_2 \:,
\end{equation}
where $\mathbf{X}\equiv(X, Y, Z)^{\textrm{T}}$ is the vector of single-qubit Pauli operators 
and $\mathbf{X}_n \equiv(X_n, Y_n, Z_n)^{\textrm{T}}$ its counterpart acting on the $N$-qubit 
Hilbert space; $\mathbbm{1}_2$ is the single-qubit identity operator.

The Lie-algebraic controllability of coupled spin-$1/2$ chains (qubit arrays) was discussed
extensively in the past~\cite{D'AlessandroBook}. In particular, for complete (operator) 
controllability -- i.e. the capability of realizing an arbitrary unitary transformation on the 
Hilbert space of the underlying system, which is tantamount to enabling universal QC -- of a 
qubit array with Ising-type interaction it is necessary to have two mutually noncommuting control 
fields acting on every qubit in the array~\cite{Wang++:16}. Therefore, an $N$-qubit system with Ising-type coupling 
between qubits and acted upon by global Zeeman-type control fields in the $x$- and $y$ directions 
[cf. Eqs.~\eqref{TotalHamiltonian}-\eqref{ControlOperators} above] is, generally speaking, not completely 
operator-controllable; rephrasing, its corresponding dynamical Lie algebra~\cite{D'AlessandroBook} 
$\mathcal{L}_{d}=\textrm{span}\{H_{ZZ},\mathcal{X},\mathcal{Y}\}$ is not isomorphic with the full 
$u(2^N)$ or $su(2^N)$ algebra, being -- in fact -- isomorphic with one of their proper Lie subalgebras. 

An important controllability-related result for Ising-coupled qubit arrays has recently been derived,
which is based on the invariance under permutations of qubits. Namely, it has been shown that a system 
described by the permutationally-invariant Hamiltonian of Eq.~\eqref{TotalHamiltonian} is completely 
state-to-state controllable for an arbitrary pair of permutationally invariant initial- and final 
states~\cite{Albertini+DAlessandro:18}. In other words, the time-dependence of global control fields 
$h_x(t)$ and $h_y(t)$ [cf. Eq.~\eqref{TotalHamiltonian}] can in principle be found such that an arbitrary 
permutationally-invariant final state can be reached in a finite time starting from a permutationally-invariant 
initial state. However -- as is also the case with other results in the realm of Lie-algebraic 
controllability~\cite{D'AlessandroBook}, which have the nature of existence theorems -- the aforementioned
recent result does not provide the actual time dependence of the fields $h_x(t)$ and $h_y(t)$ that allows 
this system to evolve from a chosen initial- to a desired final state~\cite{Zhang+Whaley:05}.

\subsection{Preparation of the Dicke state $\ket{D^{3}_{2}}$ as a quantum-control problem}  \label{DickeStatePrep}
In the following, we formulate the envisaged deterministic preparation of the two-excitation Dicke state 
$\ket{D^{3}_{2}}$ in a three-qubit system ($N=3$), starting from the state $|000\rangle$, as a quantum-control 
problem. To begin with, we provide a short reminder on the basic properties of Dicke states.

A generic $N$-qubit state with $a$ excitations (i.e. $a$ qubits in the logical state $|1\rangle$, with the 
remaining ones being in the state $|0\rangle$) can be parameterized as $\ket{\{n_1,\ldots,n_a\}}$, with 
$n_1,\ldots,n_a$ enumerating those qubits that are in the logical state $|1\rangle$. The $a$-excitation 
Dicke state of an $N$-qubit system is given by
\begin{equation}\label{DickeStateDef}
\ket{D^{N}_{a}}= \uber{N}{a}^{-1/2}\sum_{n_1<\ldots<n_a}^N
\ket{\{n_1,\ldots,n_a\}} \:,
\end{equation}
i.e. by the equal-weight superposition of all the states $\ket{\{n_1,\ldots,n_a\}}$ spanning the subspace of 
the $N$-qubit states with exactly $a$ excitations (i.e. states corresponding to bit strings of Hamming 
weight $a$); the sum in Eq.~\eqref{DickeStateDef} runs over all $\uber{N}{a}$ combinations of $a$ qubits 
out of $N$. 

While the notation used in Eq.~\eqref{DickeStateDef} is appropriate for the most general Dicke 
states, for relatively small values of $N$ one can resort to a simpler notation. In particular, the 
two-excitation Dicke state of $N=3$ qubits, the state of primary interest in the present work, can be 
written as
\begin{equation} \label{TwoExcDicke}
|D^3_2\rangle=\frac{1}{\sqrt{3}}\:(|110\rangle
+|101\rangle+|011\rangle) \:.
\end{equation}
Our treatment in what follows will rely heavily on the permutational invariance of this 
state (see Sec.~\ref{SymmSector} below).

In the special case $N=3$ the Ising-interaction Hamiltonian [cf. Eq.~\eqref{LRIsingInt}] reduces to 
\begin{equation} \label{threeQubitIsing}
H_{ZZ} = J(Z_1 Z_2+Z_2 Z_3+Z_1 Z_3)  \:,
\end{equation}
while the total control Hamiltonian is given by
\begin{eqnarray} \label{threeQubitControl}
H_C(t) &=& h_x(t)(X_1+X_2+X_3) \nonumber \\
&+& h_y(t)(Y_1 +Y_2+Y_3) \:.
\end{eqnarray}
Our objective in the following is to find the time-dependence of control fields $h_x(t)$
and $h_y(t)$ such that the dynamics governed by the total Hamiltonian $H(t)=H_{ZZ}+H_C(t)$
allows the preparation of the state $|D^3_2\rangle$ in a finite time starting from the state 
$|000\rangle$. Thus, the state $\ket{\psi(t)}$ of the three-qubit system under consideration 
ought to satisfy the conditions 
\begin{equation}\label{InitFinalStates}
\ket{\psi(t=0)} = |000\rangle \:, \quad \ket{\psi(t=T)} =\ket{D^3_2} \:,
\end{equation}
where $T$ is the as-yet-unknown state-preparation time. 

\subsection{Symmetric sector of the three-qubit Hilbert space and its basis} \label{SymmSector}
Before embarking on the design of the pulse sequence for implementing the desired Dicke-state preparation, it 
is pertinent to explore the consequences of the permutational invariance of the problem under consideration.
Particularly useful in this regard is the concept of the symmetric sector of the three-qubit Hilbert space.

In the quantum-control context, it is often beneficial to consider pure states that are invariant with 
respect to permutations of qubits~\cite{Zanardi:99,Ribeiro+Mosseri:11,Burchardt+:21}, either under 
the full symmetric group $S_N$ (where $N$ is the number of qubits) or with respect to proper subgroups 
of $S_N$~\cite{Lyons+:22}. In particular,
in the state-preparation problem under consideration we focus on the subset of all the unitaries on the 
Hilbert space $\mathcal{H}\equiv(\mathbbm{C}^2)^{\otimes 3}$ of the system at hand that are invariant under 
the full permutation group $S_3$. The relevant (permutationally invariant) Lie subgroup $U^{\textrm{S}_3}(8)$ 
of $U(8)$ has dimension $20$~\cite{Albertini+DAlessandro:18}. Its associated Lie algebra is
$u^{\textrm{S}_3}(8)=\textrm{span}\{i\varPi(\sigma_1\otimes\sigma_2\otimes\sigma_3)\}$, where 
$\varPi=(3!)^{-1}\:\sum_{P\in S_3}\:P$ and $\sigma_n$ ($n=1,2,3$) stands either for $\mathbbm{1}_{2}$ 
or one of the Pauli operators. 

The eight-dimensional total Hilbert space $\mathcal{H}$ of a three-qubit system decomposes into three invariant 
subspaces under the action of $u^{\textrm{S}_3}(8)$. Among those subspaces, which correspond to irreducible 
representations of $su(2)$, there are two subspaces of dimension $2$ and one that has dimension $4$. The latter 
comprises the states that are invariant under an arbitrary permutation of qubits, hence being referred to as the 
symmetric sector~\cite{Ribeiro+Mosseri:11}. The states 
\begin{eqnarray} \label{SSbasis}
|\zeta_0\rangle = |000\rangle \:,\quad |\zeta_1\rangle=\frac{1}{\sqrt{3}}\:
(|100\rangle+|010\rangle+|001\rangle) \:,\\
|\zeta_2\rangle=\frac{1}{\sqrt{3}}\:(|110\rangle+|101\rangle+|011\rangle) 
\:,\quad |\zeta_3\rangle = |111\rangle  \:, \nonumber
\end{eqnarray}
form an orthonormal, symmetry-adapted basis of the symmetric sector~\cite{Albertini+DAlessandro:18,Stojanovic+Nauth:22}, 
where the subscript $a$ in $|\zeta_a\rangle$ is equal to the Hamming weight of the corresponding bit string. 
These four states are, in fact, the four Dicke states $|D^3_a\rangle$ ($a=0,\ldots,3$).

The initial- and final states of our envisioned state-preparation scheme [cf. Eq.~\eqref{InitFinalStates}]
correspond to two of the basis states in Eq.~\eqref{SSbasis}. While the initial state $|000\rangle\equiv$
corresponds to $|\zeta_0\rangle\equiv|D^3_0\rangle$, the Dicke state $|D^3_2\rangle$ coincides with $|\zeta_2\rangle$. 
Therefore, it is pertinent to investigate the state-preparation problem at hand within the symmetric sector. 
To this end, we first map the four basis states in Eq.~\eqref{SSbasis} onto column vectors~\cite{Stojanovic+Nauth:22}:
\begin{eqnarray} \label{BasisColumnVecs}
\ket{\zeta_0}&\mapsto&
\left(\begin{array}{c}
1\\
0\\
0\\
0
\end{array}\right),~~\ket{\zeta_1} ~\mapsto~
\left(\begin{array}{c}
0\\
1\\
0\\
0
\end{array}\right),\nonumber\\
\ket{\zeta_2} &\mapsto&
\left(\begin{array}{c}
0\\
0\\
1\\
0
\end{array}\right),~~\ket{\zeta_3}~\mapsto~
\left(\begin{array}{c}
0\\
0\\
0\\
1
\end{array}\right).
\label{BasisinSS}
\end{eqnarray}

\section{Pulse sequence for Dicke-state preparation} \label{DickeSequence}
In the following, we first describe the construction of the NMR-type pulse sequence that allows 
one to efficiently prepare the desired Dicke state $\ket{D^{3}_{2}}$ starting from the state 
$|000\rangle$ and present the optimal values of its characteristic parameters (Sec.~\ref{PulseSequence}). 
We then discuss the feasibility of realizing the proposed state-preparation scheme
with neutral-atom qubits (based on Rydberg states~\cite{GallagherBOOK}), even in the presence of 
motion-induced dephasing and ionization effects (Sec.~\ref{NeutralAtomRealize}). Finally, we provide a 
geometrical interpretation of this pulse sequence based on a dimensional reduction of the problem to a 
two-dimensional subspace of the original three-qubit Hilbert space and the concept of the 
Bloch sphere (Sec.~\ref{GeomInterpret}).

As indicated above, the sought-after Dicke state is invariant with respect to permutations of qubits. 
Therefore, the state-preparation problem at hand can be reduced to a four-dimensional subspace (symmetric 
sector) of the three-qubit Hilbert space and formulated using the basis given in Eq.~\eqref{SSbasis}. 
We hereafter also set $\hbar=1$, thus all the timescales in the problem under consideration will be 
expressed in terms of the inverse Ising-coupling strength $J^{-1}$.

\subsection{Layout of the pulse sequence and the optimal parameter values}  \label{PulseSequence}
\begin{figure}[t!]
\includegraphics[width=0.9\linewidth]{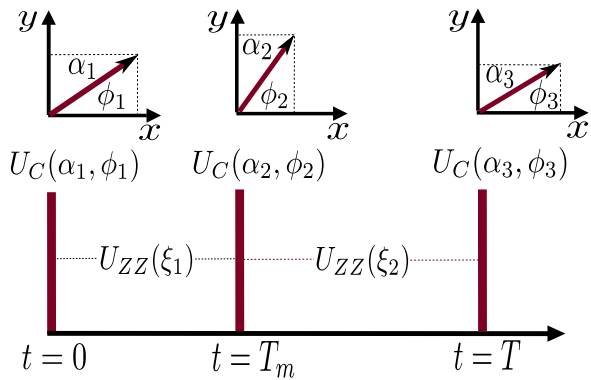}
\caption{\label{fig:PulseSequence}(Color online)Schematic of the pulse sequence for the 
preparation of the Dicke state $\ket{D^{3}_{2}}$, which consists of three instantaneous global 
control pulses and two Ising-interaction pulses. The control pulse characterized by the vector 
$\boldsymbol{\alpha}_j$ ($j=1,2,3$) in the $x$-$y$ plane corresponds to a global qubit rotation through 
an angle of $2\alpha_j$ around the axis whose direction is determined by the unit vector $\mathbf{\hat{n}}_{j}
\equiv(\cos\phi_j,\sin\phi_j,0)^{\textrm{T}}$; the polar angles $\phi_j$ are restricted to values 
in $[0,\pi)$, while positive (negative) values of $\alpha_j$ correspond to counterclockwise
(clockwise) rotations. Here $U_C(\boldsymbol{\alpha}_j)\equiv U_C(\alpha_j,\phi_j)$ are the time-evolution 
operators corresponding to these control pulses; $U_{ZZ}(\xi_1)$ and $U_{ZZ}(\xi_2)$ are their 
counterparts pertaining to the interaction pulses, with $\xi_1\equiv JT_m$ and $\xi_2\equiv J(T-T_m)$ 
being their respective dimensionless durations.}
\end{figure}

In what follows, we aim to find a solution of the state-preparation problem under consideration [cf. Eq.~\eqref{InitFinalStates}]
that has the form of an NMR-type pulse sequence; such pulse sequences typically comprise a certain number of 
instantaneous ($\delta$-shaped) global control pulses and interaction pulses between consecutive control 
pulses. More specifically yet, we assume that the pulse sequence of interest here consists of three such 
control pulses in transverse directions -- at times $t=0$, $t=T_m$, and $t=T$ -- and two Ising-interaction 
pulses (a pictorial illustration of this pulse sequence is given in Fig.~\ref{fig:PulseSequence}). 
The transverse (global) control field $\boldsymbol{h}(t)\equiv[h_x(t), h_y(t),0]^{\textrm{T}}$ 
can then be written in the form
\begin{equation}\label{eq:h(t)_def}
\boldsymbol{h}(t)=\boldsymbol{\alpha}_1\delta(t)
+\boldsymbol{\alpha}_2\delta(t-T_m) 
+ \boldsymbol{\alpha}_3\delta(t-T)\:,
\end{equation}
where the presence of the delta functions in the last equation captures the instantaneous character of the 
envisioned control pulses and $\boldsymbol{\alpha}_1$, $\boldsymbol{\alpha}_2$, $\boldsymbol{\alpha}_3$ 
are, in principle, arbitrary vectors in the $x$-$y$ plane. The directions of these vectors are specified 
by their respective polar angles (azimuths) $\phi_1$, $\phi_2$, $\phi_3$. 

To understand the connection between instantaneous global control pulses in the system at hand 
and global rotations it is useful to recall the well-known identity  
\begin{equation}\label{PauliExpIdentity}
\exp[-i\theta (\mathbf{\hat{n}}\cdot \mathbf{X})]= 
\cos\theta \mathbbm{1}_{2}- i\sin\theta\:(\mathbf{\hat{n}}\cdot \mathbf{X})  \:,
\end{equation}
for the single-qubit rotation operators $R_{\mathbf{\hat{n}}}(2\theta)\equiv\exp[-i\theta (\mathbf{\hat{n}}\cdot\mathbf{X})]$,
where $\mathbf{\hat{n}}$ is an arbitrary unit vector. Based on the last identity it is straightforward to infer that 
an instantaneous control pulse described by the vector $\boldsymbol{\alpha}_j$ ($j=1,2,3$) corresponds to a global 
rotation through an angle of $2\alpha_j$ around the axis whose direction is determined by the unit vector $\mathbf{\hat{n}}_{j}
\equiv(\cos\phi_j,\sin\phi_j,0)^{\textrm{T}}$. We choose the convention according to which the angles $\phi_j$ are 
restricted to values in $[0,\pi)$, while positive (negative) values of $\alpha_j$ correspond to counterclockwise
(clockwise) rotations. 

It is appropriate to comment at this point on the feasibility of realizing pulse sequences
of the proposed form, which involve global instantaneous control pulses, in different physical 
platforms for QC. Firstly, it should be stressed that the assumption about the instantaneous 
character of control pulses is well justified in systems where the magnitude of control fields 
is much larger than the qubit-qubit coupling strength. This last requirement 
is satisfied for typical control fields of magnetic origin used in the NMR domain~\cite{Vandersypen+Chuang:05}, 
as well as for control fields used in superconducting-qubit-~\cite{Geller+:10} and neutral-atom 
systems~\cite{Morgado+Whitlock:21}. Secondly, the global character of control pulses -- leading 
to global qubit rotations -- constitutes a necessity in several QC platforms of current interest. 
A typical example are neutral-atom QC setups, where the role of two relevant logical qubit states 
is played either by two hyperfine states or by a ground state and a high-lying Rydberg state. In 
such systems one typically makes use of a global microwave field to perform a rotation in the $x-y$ 
plane on every qubit. This rotation gate has to be global in character because 
the distance between qubits in such systems is several orders of magnitude smaller than the wavelength 
of the microwave field. 

The explicit expressions for the time-evolution operators $U_C(\boldsymbol{\alpha}_j)\equiv U_C(\alpha_j,\phi_j)$ 
($j=1,2,3$) corresponding to the three global control pulses, as well as their Ising-interaction counterparts 
$U_{ZZ}(\xi_i)$ with respective dimensionless (defined in units of $J^{-1}$) 
durations $\xi_i$ ($i=1,2$), in the symmetry-adapted basis of Eq.~\eqref{SSbasis} are straightforward to 
derive [for a detailed derivation, see Appendix~\ref{DerivTimeEvol}]. By making use of those expressions, 
we can recover the full time-evolution operator 
\begin{eqnarray} \label{fullU}
\mathcal{U}_{\textrm{PS}}(\xi_1,\xi_2,\boldsymbol{\alpha}_1,\boldsymbol{\alpha}_2,
\boldsymbol{\alpha}_3) &=& U_C(\boldsymbol{\alpha}_3)U_{ZZ}(\xi_2)U_C(\boldsymbol{\alpha}_2) \nonumber \\
&\times& U_{ZZ}(\xi_1)U_C(\boldsymbol{\alpha}_1)
\end{eqnarray}
describing the envisioned pulse sequence (for a pictorial illustration, see Fig.~\ref{fig:PulseSequence}).

Aiming to prepare the state $\ket{D^3_2}$ starting from the initial state $\ket{000}$, 
we maximize the figure of merit of relevance here -- the Dicke-state fidelity 
$\mathcal{F}_{\textrm{D}}$ -- with respect to the eight pulse-sequence parameters ($\xi_1$, 
$\xi_2$, $\alpha_1$, $\alpha_2$, $\alpha_3$, $\phi_1$, $\phi_2$, $\phi_3$). 
This fidelity is given by the module of the overlap of the target state $\ket{D^3_2}$ and the 
actual final state $\,\mathcal{U}_{\textrm{PS}}(\xi_1,\xi_2,\boldsymbol{\alpha}_1,\boldsymbol
{\alpha}_2,\boldsymbol{\alpha}_3)\ket{000}$ of the three-qubit system that results from the 
application of the pulse sequence [cf. Eq.~\eqref{fullU}]:
\begin{equation} \label{DickeFidelity}
\mathcal{F}_{\textrm{D}}=\left|\bra{D^3_2}\,\mathcal{U}_{\textrm{PS}}(\xi_1,\xi_2,
\boldsymbol{\alpha}_1,\boldsymbol{\alpha}_2,\boldsymbol{\alpha}_3)\ket{000}\right| \:.
\end{equation}

We numerically optimize the Dicke-state fidelity in Eq.~\eqref{DickeFidelity} by making use of 
the \texttt{minimize} routine from the \texttt{scipy.optimization} package of the SciPy 
library~\cite{minimize_scipy}. In this way, we obtain the following optimal values of the 
eight pulse-sequence parameters:
\begin{eqnarray} \label{OptValuesParams}
\alpha_{1,0} &=& 3\pi/4  \:, \nonumber \\
\alpha_{2,0} &=& -\arccos(1/3)/4 \:, \nonumber\\
\alpha_{3,0} &=& \pi/4   \:,  \\
\phi_{1,0} &=& \phi_{3,0} = \pi/2  \:, \quad \phi_{2,0} = 0\:, \nonumber \\
\xi_{1,0} &=& \xi_{2,0} = [\pi - \arccos(1/3)]/4 \:. \nonumber 
\end{eqnarray}
Based on the obtained results, we can specify the $x$- and $y$ components of the global 
control field $\boldsymbol{h}(t)$ [whose assumed general form is given in Eq.~\eqref{eq:h(t)_def}]
that are required for the preparation of the state $\ket{D^3_2}$ in the system at hand:
\begin{eqnarray}\label{}
h_x (t) &=& -\frac{J}{4}\:\arccos(1/3) \:\delta(t-T_m)  \:,\nonumber \\
h_y (t) &=& \frac{\pi J}{4}\:[3\:\delta(t)+\delta(t-T)]\:.
\end{eqnarray}
Thus, the $x$ component of the obtained global control field has the form of an instantaneous 
pulse at $t=T_m$, while its counterpart in the $y$ direction entails two such pulses, at times 
$t=0$ and $t=T$.

From the obtained optimal parameter values [cf. Eq.~\eqref{OptValuesParams}], the following conclusions can be 
drawn. Firstly, the first- and third control pulses correspond to global rotations around the $y$ axis, with the 
respective rotation angles of $3\pi/2$ and $\pi/2$. Secondly, the second control pulse corresponds to a global 
rotation around the $x$ axis through an angle of $\arccos(1/3)/2$. Finally, the two interaction pulses 
have equal durations. Given that the pulse sequence involves instantaneous control pulses, the total duration 
$T$ of the pulse sequence -- i.e. the state-preparation time -- is given by $T=2\xi_{1,0}\:J^{-1}$, which 
amounts to $T=[\pi-\arccos(1/3)]J^{-1}/2$. Upon reinstating $\hbar$, the total Dicke-state preparation time 
within the framework of the proposed scheme is thus given by
\begin{equation} \label{PSduration}
T\approx 0.95\:\frac{\hbar}{J} \:.
\end{equation}

\subsection{Realization with neutral-atom qubits} \label{NeutralAtomRealize}
In order to examine the practical feasibility of this scheme, it is of interest to estimate the order of magnitude 
of the obtained state-preparation time in QC platforms of current interest. For instance, one promising platform 
is based on optically-trapped neutral atoms confined in optical tweezers at a typical mutual distance of a few
micrometers~\cite{Morgado+Whitlock:21,Kaufman+Ni:21}. In such systems it is nowadays possible to prepare almost any 
two-dimensional arrangement of neutral atoms; as a special case, one can prepare a system that 
consists of three equidistant neutral atoms. Assuming that the role of the logical qubit states $|0\rangle$ and 
$|1\rangle$ is played by the atomic ground state $|g\rangle$ and a high-lying Rydberg state $|r\rangle$~\cite{GallagherBOOK}, 
respectively, such a system of three neutral-atom qubits is of direct relevance for the present work.

The Ising-type interaction between neutral-atom qubits originates from the van-der-Waals (vdW) type interaction 
between atoms. This interaction is given by 
\begin{eqnarray}
V_{\textrm{vdW}} &=& \sum_{n<n'}\frac{C_6}{R^6_{nn'}}\:|r_n r_{n'}
\rangle\langle r_n r_{n'}| \nonumber\\
&=& \frac{1}{4}\:\sum_{n<n'}\frac{C_6}{R^6_{nn'}}\:(Z_n Z_{n'} 
+ Z_n + Z_{n'} + 1)\:,
\end{eqnarray}
where $R_{nn'}$ is the distance between atoms $n$ and $n'$ and $C_6$ the vdW interaction constant. Thus, the 
Ising-coupling strength between Rydberg-atom qubits $n$ and $n'$ is given in this system by $J_{nn'}=C_6/(4R^6_{nn'})$.
For a typical interatomic distance of around $5\:\mu$m and a value around $50$ for the principal quantum number~\cite{Morgado+Whitlock:21} 
we have $J/\hbar \gtrsim 1.5$\:MHz. Therefore, based on Eq.~\eqref{PSduration} we find that $T\lesssim 0.6\:\mu$s, 
which squares with the expectation that the time needed to carry out a typical entangling operation in such systems 
should be of the order of $1\:\mu$s. The obtained state-preparation time is thus at least two orders of magnitude 
shorter than the typical radiative lifetimes of Rydberg states, which for the chosen range of principal quantum numbers 
are of the order of $100\:\mu$s. This speaks in favor of the feasibility of the proposed scheme for the 
preparation of the state $\ket{D^3_2}$.

It is of interest to examine the feasibility of realizing the proposed state-preparation scheme 
with neutral-atom qubits in the presence of potentially debilitating effects such as motion-induced dephasing and
ionization. In what follows, we demonstrate that under typical experimental conditions -- such as the Lamb-Dicke 
regime of atomic motion and the presence of a cryogenic environment -- these effects do not pose obstacles to the 
realization of the desired Dicke state in a system of neutral atoms confined in optical dipole traps 
(tweezers)~\cite{Kaufman+Ni:21}.

Generally speaking, laser-based manipulation of atomic states gives rise 
to undesired phases that depend on atomic positions. These effects are, however, alleviated 
when atoms are initially confined in the ground state of the trapping potential (in this 
particular case, that of an optical tweezer) and the system is in the Lamb-Dicke regime. 
The latter refers to situations where the trap is sufficiently confining that the momentum
imparted by the scattered photon does not cause a change in the motional state of an 
atom~\cite{Kaufman+Ni:21}. In other words, the recoil energy of an atom is much smaller
than the spacing between adjacent vibrational levels corresponding to the trapping potential.

Let us consider an atom of mass $m$ whose position within a 
tweezer trap with frequency $\omega_{\textrm{tr}}$ is given by $x_{n}=x^{0}_{n}+s_{n}$, 
where $x^{0}_{n}$ is its equilibrium position and 
$s_{n}=l_0(a_n+a^{\dagger}_n)$ the fluctuation due to its motion; here $l_0 \equiv\sqrt{\hbar/
(2m\omega_{\textrm{tr}})}$ is the harmonic zero-point length corresponding to the motional 
ground state within a tweezer trap, while the creation (annihilation) operator 
$a^{\dagger}_n$ ($a_n$) creates (destroys) a single excitation pertaining to the motional
degrees of freedom of the considered atom. The atom is assumed to initially be in the state
$|g\rangle_n|0\rangle_n \equiv |g,0\rangle_n$, i.e. in the atomic ground state $|g\rangle$ 
and in the ground state $|0\rangle$ with respect to its motional degrees of freedom. The 
action of the Hamiltonian describing an atom-laser interaction on this state leads to the
state $\Omega(t)e^{ikx_{n}}\:|e,0\rangle_n$, where $\Omega(t)$ is the Rabi frequency 
of the external laser and $|e\rangle_n$ is an excited atomic state (one special case of 
such states is the Rydberg state $|r\rangle_n$). In the Lamb-Dicke regime, defined by 
the condition $\eta \equiv kl_0/\sqrt{2} \ll 1$, we can expand the exponential factor 
$e^{ikx_{n}}\equiv e^{ikx^{0}_{n}} e^{iks_{n}}$ and thereby obtain
\begin{equation} \label{LDexp}
e^{ikx_{n}} = e^{ikx^{0}_{n}}[1+i\eta (a_n+a^{\dagger}_n)
+ \mathcal{O}(\eta^2)] \:.
\end{equation}
By first absorbing the constant prefactor $e^{ikx^{0}_{n}}$ into the definition of the atomic
basis states, i.e. defining $|\tilde{e}\rangle_n\equiv e^{ikx^{0}_{n}}|e\rangle_n$, the Hamiltonian 
describing the laser-induced excitation of atom $n$ in the basis $\{|g,0\rangle_n,|\tilde{e},0\rangle_n,
|\tilde{e},1\rangle_n\}$ adopts the form
\begin{equation} \label{Hexc}
H_{\textrm{exc}} = \begin{pmatrix}
0 & \Omega(t) & \eta\Omega(t) \\
\Omega(t) & 0 & 0 \\
\eta\Omega(t) & 0 & \omega_{\textrm{tr}}
\end{pmatrix} \:.
\end{equation}
If the considered atom starts in the state $|g,0\rangle_n$, the probability of populating 
the state $|\tilde{r},1\rangle_n$ (a special case of $|\tilde{e},1\rangle_n$), which corresponds 
to the situation where the excitation of the considered atom to the desired Rydberg state is 
accompanied by the creation of a single motional quantum within the tweezer trap, will be negligible 
if $\Omega\eta \ll \omega_{\textrm{tr}}$. For a typical choice of parameter values of our envisioned
system [$\Omega\sim 1$\:kHz, $\omega_{\textrm{tr}}/(2\pi)\sim (1 - 10)$\:kHz] being in the Lamb-Dicke 
regime ($\eta\ll 1$) immediately implies that this last condition is fulfilled. Therefore, the 
motion-induced dephasing does not have an appreciable detrimental impact on the realization of our 
state-preparation scheme.

Another possible source of decoherence in our envisioned neutral-atom system is 
depopulation of the Rydberg state. The rate of Rydberg-state depopulation is given by 
the sum of probabilities of spontaneous emission to all lower states, which are given by
Einstein's coefficients $A_{if}$~\cite{GallagherBOOK} (with $|i\rangle$ and $|f\rangle$ 
being the initial and final states, respectively):
\begin{equation} \label{DepopRate}
\tau_r^{-1} = \sum_f A_{if} = \frac{2e^2}{3\epsilon_0 c^3 h}
\:\sum_{E_f<E_i}\omega^{3}_{if}\:|\langle i|\mathbf{r}|f\rangle|^2 \:.
\end{equation} 
Here $\omega_{if}\equiv (E_i - E_f)/\hbar$ is the transition frequency 
and $\langle i|\mathbf{r}|f\rangle$ the dipole matrix element between 
initial and final states; importantly, the sum in Eq.~\eqref{DepopRate} runs 
only over final states with energies smaller than that of the initial state. 
One of the possible Rydberg-decay channels corresponds to loss due to atomic 
collisions; however, such loss is heavily suppressed in the envisioned system 
because it is assumed that in this system there is only one atom in each individual 
tweezer trap. The other possible Rydberg-decay channel pertains to blackbody-radiation-induced 
transitions~\cite{GallagherBOOK}. Yet, in a cryogenic environment, which the
proposed system is assumed to operate in, such transitions are known to be negligible.

\subsection{Geometric interpretation of the pulse sequence on the Bloch sphere} \label{GeomInterpret}
Making use of the permutational-symmetry of the state-engineering problem under consideration
allowed us to treat this problem in the four-dimensional symmetric sector [recall Sec.~\ref{SymmSector}]
rather than in the original, eight-dimensional Hilbert space of the three-qubit system. Further
dimensional reduction of the problem at hand -- to a two-dimensional space -- is enabled through 
the decomposition of the relevant operators in the (two-dimensional) eigensubspaces of the $x$- 
and $y$ parity operators; the latter are given by the tensor products of Pauli $X$ and $Y$ operators 
on different qubits:
\begin{equation}\label{ParityOper}
X_p\equiv X\otimes X \otimes X \:,\quad Y_p\equiv Y\otimes Y \otimes Y \:.
\end{equation}
In what follows, we make use of this additional dimensional reduction in order to provide a 
geometric interpretation of the proposed pulse sequence on the Bloch sphere.

It is worthwhile to first note that the operators $H_{ZZ}$ and $\mathcal{X}=X_1+X_2+X_3$ commute with 
the $x$-parity operator $X_p$ (similarly, the operators $H_{ZZ}$ and $\mathcal{Y}=Y_1+Y_2+Y_3$ commute 
with the $y$-parity operator $Y_p$). As a result, each of the two (degenerate) eigensubspaces of $X_p$, 
which correspond to the eigenvalues $\pm 1$, are invariant with respect to the operators $H_{ZZ}$ 
and $\mathcal{X}$. This allows one to simplify the analysis of the dynamics inherent to $H_{ZZ}$ 
and $\mathcal{X}$ by projecting these operators to the $+1$ and $-1$ eigensubspaces of $X_p$.

While, generally speaking, the dimensionality of the $+1$ and $-1$ eigensubspaces of $X_p$ is four, 
in the problem at hand we are only interested in their corresponding eigenstates that belong to the 
symmetric sector (recall the discussion in Sec.~\ref{SymmSector}). Thus, the relevant two-dimensional 
$+1$ eigensubspace of $X_p$ is spanned by the states that in our chosen symmetric-sector basis [cf. 
Eq.~\eqref{SSbasis}] are given by
\begin{eqnarray} 
\ket{v_1} = \frac{1}{\sqrt{2}}
\left(\begin{array}{c}
1\\
0\\
0\\
1
\end{array}\right),~~\ket{v_2} = \frac{1}{\sqrt{2}}
\left(\begin{array}{c}
0\\
1\\
1\\
0
\end{array}\right)\:, \label{plus1states}
\end{eqnarray}
while its $-1$ counterpart is spanned by
\begin{eqnarray} \label{minussub}
\ket{v_3} =  \frac{1}{\sqrt{2}}
\left(\begin{array}{c}
1\\
0\\
0\\
-1
\end{array}\right)\:,~~\ket{v_4} = \frac{1}{\sqrt{2}}
\left(\begin{array}{c}
0\\
1\\
-1\\
0
\end{array}\right) \:. \label{minus1states}
\end{eqnarray}

In the $-1$ eigensubspace of $X_p$ the actions of $H_{ZZ}$ and $\mathcal{X}$
are represented, respectively, by the matrices
\begin{equation}
H_{ZZ}^{(-1)}/J = \begin{pmatrix}3 & 0\\ 
0 & -1\end{pmatrix} \:, ~~\mathcal{X}^{(-1)} = \begin{pmatrix}0 & \sqrt{3}\\
\sqrt{3} & -2\end{pmatrix} \:.\label{reducePlus1}
\end{equation}
It is straightforward to express these $2\times 2$ matrices in the basis 
$\{\mathbbm{1}_{2}, X, Y, Z\}$
\begin{eqnarray}
H_{ZZ}^{(-1)}/J &=& \mathbbm{1}_{2}+2Z \:, \\
\mathcal{X}^{(-1)} &=& -\mathbbm{1}_{2}+\sqrt{3}\:X+Z \nonumber\:,
\end{eqnarray}
which can further be recast in the form
\begin{eqnarray}\label{HzzKappaMinSub}
H_{ZZ}^{(-1)}/J &=& \mathbbm{1}_{2}+2(0,0,1)^{\textrm{T}}\cdot \mathbf{X} \:, \\
\mathcal{X}^{(-1)} &=& -\mathbbm{1}_{2}+2(\sqrt{3}/2,0,1/2)^{\textrm{T}}\cdot 
\mathbf{X} \nonumber\:.
\end{eqnarray}
The form of the last equations implies that the time-evolution operators corresponding 
to the operators $H_{ZZ}^{(-1)}$ and $\mathcal{X}^{(-1)}$, acting in the $-1$ eigensubspace 
of $X_p$, have the form of global-rotation operators. Their respective rotation axes are 
defined by the unit vectors $(0,0,1)^{\textrm{T}}$ and $(\sqrt{3}/2,0,1/2)^{\textrm{T}}$.

The adopted initial state $|000\rangle$ of our state-preparation scheme is an 
equal-weight superposition of states in the $+1$ and $-1$ eigensubspaces of the
parity operator, more precisely $|000\rangle=(|v_1\rangle + |v_3\rangle)/\sqrt{2}$.
At the same time, the target state $\ket{D^3_2}$ is given by $\ket{D^3_2}=(|v_2\rangle
-|v_4\rangle)/\sqrt{2}$.

The proposed five-stage pulse sequence, which gives rise to the state $\ket{D^3_2}$ starting 
from $|000\rangle$, consists of three global-rotation- and two Ising-interaction pulses, i.e.
\begin{eqnarray} \label{PSeqn}
\ket{D^3_2} &=& U_C(\alpha_{3,0},\phi_{3,0})U_{ZZ}(\xi_{2,0})U_C(\alpha_{2,0},\phi_{2,0}) \nonumber\\
&\times& U_{ZZ}(\xi_{1,0})U_C(\alpha_{1,0},\phi_{1,0})\:|000\rangle \:,
\end{eqnarray}
where $\{\alpha_{1,0},\alpha_{2,0},\alpha_{3,0},\phi_{1,0},\phi_{2,0},\phi_{3,0},\xi_{1,0},\xi_{2,0}\}$
are the optimal values of the pulse-sequence parameters given by Eq.~\eqref{OptValuesParams}].
The last equation can straightforwardly be manipulated to the form
\begin{equation} \label{PSeqnEquiv}
|D\rangle = U_{ZZ}(\xi_{2,0})U_C(\alpha_{2,0},\phi_{2,0})
U_{ZZ}(\xi_{1,0})\:|A\rangle \:,
\end{equation}
where $|A\rangle= U_C(\alpha_{1,0},\phi_{1,0})\:|000\rangle\equiv R_y(3\pi/2)\:|000\rangle$ is the state of 
the three-qubit system after the first global control pulse (a rotation around the $y$-axis through an angle 
of $3\pi/2$) is carried out, while $|D\rangle = U^{-1}_C(\alpha_{3,0},\phi_{3,0})\:\ket{D^3_2}\equiv R_y(-\pi/2)
\:\ket{D^3_2}$ is its state before the third control pulse (a rotation around the $y$-axis through an angle 
of $\pi/2$) is performed. In terms of the basis vectors $|v_3\rangle$ and $|v_4\rangle$ of the $-1$ eigensubspace 
of the $x$-parity operator $X_p$ [cf. Eq.~\eqref{minussub}], the states $|A\rangle$ and $|D\rangle$ are given 
by $|A\rangle=(\sqrt{3}|v_4\rangle-|v_3\rangle)/2$ and $|D\rangle=(\sqrt{3}|v_3\rangle+|v_4\rangle)/2$.

\begin{figure}[t!]
\includegraphics[width=0.9\linewidth]{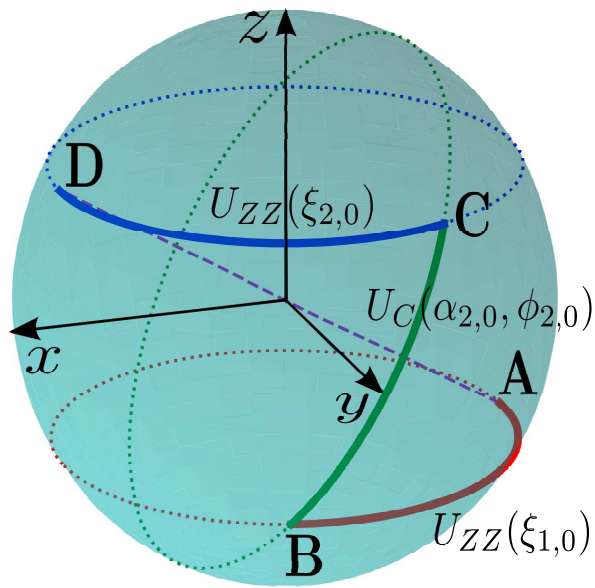}
\caption{\label{fig:PSonBlochSphere}(Color online)Geometric interpretation of the proposed pulse sequence
on the Bloch sphere, which pertains to the two-dimensional eigensubspace of the $x$-parity operator $X_p$ 
that corresponds to the eigenvalue $-1$. The point A on the Bloch sphere corresponds to the state $|A\rangle
\equiv R_y(3\pi/2)\:|000\rangle$ of the system 
after the first global control pulse, while D pertains to its state $|D\rangle\equiv R_y(-\pi/2)\:\ket{D^3_2}$ 
before the third control pulse; the dashed line indicates the global-rotation axis corresponding to the 
operator $\mathcal{X}^{(-1)}$ in Eq.~\eqref{HzzKappaMinSub}.}
\end{figure}

While the state $|000\rangle$ -- as explained above -- represents a linear combination of states from two 
different eigensubspaces of $X_p$, the state  $|A\rangle$ obtained from the latter through the rotation 
$R_y(3\pi/2)$ belongs to the $-1$ eigensubspace. Because the same is true of the state $|D\rangle$, 
all three remaining stages of the pulse sequence -- represented by the operators 
$U_{ZZ}(\xi_{1,0})$, $U_C(\alpha_{2,0},\phi_{2,0})$, and $U_{ZZ}(\xi_{2,0})$ in Eq.~\eqref{PSeqnEquiv} -- lead 
to the evolution of the three-qubit system within that same (two-dimensional) eigensubspace of $X_p$. 
Given that it is confined to a two-dimensional subspace, this evolution can be visually represented 
using a Bloch sphere. 

Before even attempting to provide a geometrical interpretation of the proposed pulse sequence, it is useful 
to recall that an arbitrary pure single-qubit state can be parametrized as~\cite{NielsenChuangBook}
\begin{equation}
|\Psi(\theta,\phi)\rangle = \cos\frac{\theta}{2}\:|0\rangle 
+ e^{i\phi}\sin\frac{\theta}{2}\:|1\rangle \:,
\end{equation}
where the values of the polar angle $\theta\in [0,\pi]$ and the azimuthal angle $\phi\in [0,2\pi)$ 
define the point $(x,y,z)=(\sin\theta\cos\phi,\sin\theta\sin\phi,\cos\theta)$ that corresponds to 
the state $|\Psi(\theta,\phi)\rangle$ on the Bloch sphere. According to this parametrization, the 
north pole of the Bloch sphere ($\theta=0$) corresponds to the logical $|0\rangle$ state, while 
the south pole ($\theta=\pi$) represents the state $|1\rangle$. 

For our present purpose of geometrically representing unitary transformations in the relevant 
two-dimensional subspace of the original three-qubit Hilbert space, we choose the convention 
in which the eigenvector $|v_3\rangle$ is identified with the north pole of the Bloch sphere
in Fig.~\ref{fig:PSonBlochSphere} and $|v_4\rangle$ with the south pole [cf. Eq.~\eqref{minussub}]. 
On this Bloch sphere the states $|A\rangle$ and $|D\rangle$ are represented by the eponymous 
points. At the same time, the three stages of the proposed pulse sequence that are required 
to steer the system from $A$ to $D$ are represented by following three curves: the curve 
A - B [the first Ising-interaction pulse, described by $U_{ZZ}(\xi_{1,0})$], B - C 
[the second control pulse, i.e. a rotation around the $x$-axis through an angle of $-\arccos(1/3)/2$, 
represented by $U_C(\alpha_{2,0},\phi_{2,0})$], and C - D [the second Ising-interaction pulse; its 
corresponding time-evolution operator is $U_{ZZ}(\xi_{2,0})$]. All three curves have the same shape,
because in the relevant two-dimensional subspace of the original three-qubit Hilbert space both control
pulses and their Ising-interaction counterparts act as global-rotation operators, as implied by the 
form of Eq.~\eqref{HzzKappaMinSub}.

\section{Robustness of the state-preparation scheme} \label{PulseSequenceRobust}
Having found the idealized pulse sequence for the preparation of the desired Dicke state [cf. Eq.~\eqref{OptValuesParams} 
in Sec.~\ref{PulseSequence}], it is pertinent to quantitatively assess the sensitivity of the proposed state-preparation
scheme to various imperfections. Typical imperfections considered in NMR-type pulse sequences~\cite{Vandersypen+Chuang:05} 
pertain to errors in the rotation axes (more precisely, in the directions of their corresponding unit vectors $\mathbf{\hat{n}}$) 
and/or errors in the rotation angles. Thus, the actual qubit rotation carried out experimentally is not the ideal one represented 
by $R_{\mathbf{\hat{n}}}(2\theta)\equiv\exp[-i\theta (\mathbf{\hat{n}}\cdot\mathbf{X})]$ [cf. Eq.~\eqref{PauliExpIdentity}].
A realistic rotation is instead described by
\begin{equation}
\tilde{R}_{\mathbf{\hat{n}}}(2\theta)=\exp\left[-i\:\mathbf{f}
(\theta,\mathbf{\hat{n}})\cdot\mathbf{X}\right]\:,
\end{equation}
where the vector function $\mathbf{f}(\theta,\mathbf{\hat{n}})$ characterizes various types of systematic errors~\cite{Vandersypen+Chuang:05}.
For example, the form $\mathbf{f}(\theta,\mathbf{\hat{n}})=\theta(1+\varepsilon_{\theta})\mathbf{\hat{n}}$ of this vector function 
allows one to describe under- and over-rotation errors (respectively for negative- and positive values of $\varepsilon_{\theta}$). 
On the other hand, by choosing this function to have the form $\mathbf{f}(\theta,\mathbf{\hat{n}})=\theta(n_x\cos\varepsilon_{\phi}
+n_y\sin\varepsilon_{\phi}, n_y\cos\varepsilon_{\phi}-n_x\sin\varepsilon_{\phi}, n_z)^{\textrm{T}}$ one can capture the error in 
the rotation axis~\cite{Vandersypen+Chuang:05}, whose nominal direction is given by the unit vector $\mathbf{\hat{n}}\equiv(\cos\phi,
\sin\phi,0)^{\textrm{T}}$.

In keeping with the above general considerations, we investigate the robustness of the proposed scheme
by taking into account systematic errors in the global-rotation angles corresponding to the three control pulses (i.e. errors 
in the values of the parameters $\alpha_1$, $\alpha_2$, and $\alpha_3$), errors in the directions of the attendant rotation axes 
(i.e. errors in the parameters $\phi_1$, $\phi_2$, and $\phi_3$), and, finally, errors corresponding to the durations of the 
Ising-interaction pulses ($\xi_1$ and $\xi_2$). In other words, we consider errors $\varepsilon_p$ of either sign for each of 
the eight characteristic pulse-sequence parameters ($p=\xi_1,\xi_2,\alpha_1,\alpha_2,\alpha_3,\phi_1, \phi_2,\phi_3$): 
\begin{eqnarray}
\xi_i &=& \xi_{i,0}(1+\varepsilon_{\xi_i}) \quad (\:i=1,2\:) \:,\\
\alpha_j &=& \alpha_{j,0}(1+\varepsilon_{\alpha_j})\:,\quad
\phi_j=\phi_{j,0}+\varepsilon_{\phi_j} \quad (\:j=1,2,3\:)\:. \nonumber
\end{eqnarray}
In connection with the form of the last equation, it is important to stress that errors in the parameters 
$\xi_1$, $\xi_2$, $\alpha_1$, $\alpha_2$, and $\alpha_3$ have the nature of relative errors, while for 
$\phi_1$, $\phi_2$, and $\phi_3$ it is pertinent to consider absolute errors. In both cases, we assume 
the corresponding errors to vary between $-0.1$ and $0.1$. 

\begin{figure}[t!]
\includegraphics[width=0.95\linewidth]{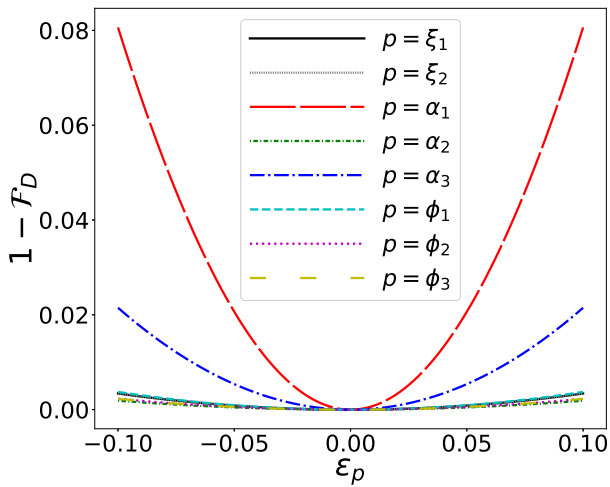}
\caption{\label{fig:InfidelityVarParams}(Color online)Dependence of the deviation of the Dicke-state 
fidelity from unity $1-\mathcal{F}_\text{D}$ on the errors $\varepsilon_p$ in the values of the parameters 
that characterize the pulse sequence for implementing the Dicke-state preparation.}
\end{figure}

The relative impacts on the target-state fidelity $\mathcal{F}_{\textrm{D}}$ of the individual deviations 
$\varepsilon_p$ from the respective optimal values of the eight relevant parameters in the problem at hand, 
are depicted in Fig.~\ref{fig:InfidelityVarParams}, which shows the deviation of the fidelity from unity (i.e. 
the infidelity) $1-\mathcal{F}_{\textrm{D}}$. What can be inferred from these results is that -- among 
these eight parameters -- the Dicke-state fidelity is by far most sensitive to deviations from the optimal 
value of the parameter $\alpha_1$, i.e. the global-rotation angle pertaining to the first ($t=0$) control 
pulse. Another parameter whose variation significantly affects the target-state fidelity is $\alpha_3$, 
which determines the global-rotation angle corresponding to the third  ($t=T$) control pulse. The errors 
in the values of the remaining six parameters have much smaller bearing on the target-state fidelity. 

\begin{figure}[b!]
\includegraphics[width=0.95\linewidth]{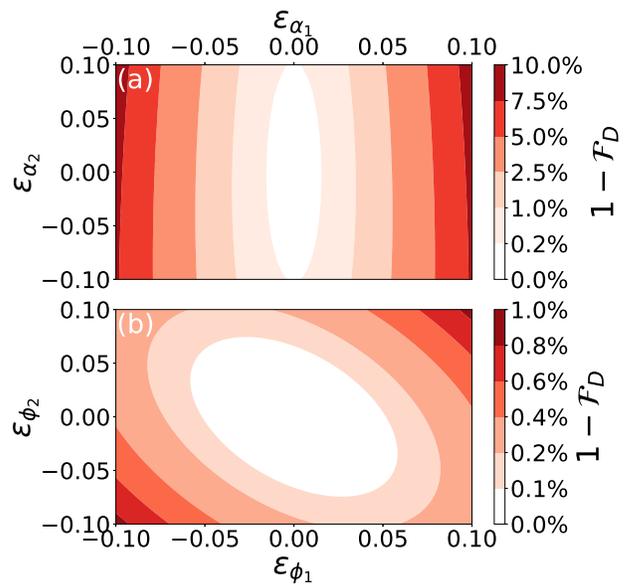}
\caption{\label{fig:InfidelityDensPlotTwoParam1}(Color online)Dependence of the deviation $1-\mathcal{F}_\text{D}$ 
of the Dicke-state fidelity from unity on (a) errors in the parameters $\alpha_1$ and $\alpha_2$,
i.e. deviations from their respective optimal values $\alpha_{1,0}=3\pi/4$ and $\alpha_{2,0}=-\arccos(1/3)/4$,
and (b) errors in the parameters $\phi_1$ and $\phi_2$, i.e. deviations from their respective optimal values 
$\phi_{1,0} = \pi/2$ and $\phi_{2,0}=0$.}
\end{figure}

\begin{figure}[t!]
\includegraphics[width=0.95\linewidth]{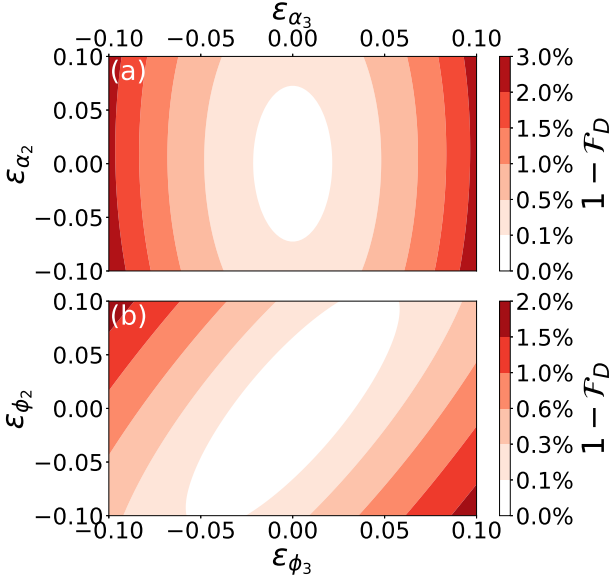}
\caption{\label{fig:InfidelityDensPlotTwoParam2}(Color online)Dependence of the deviation $1-\mathcal{F}_\text{D}$ 
of the Dicke-state fidelity from unity on (a) errors in the parameters $\alpha_2$ and $\alpha_3$, i.e. 
deviations from their respective optimal values $\alpha_{2,0}=-\arccos(1/3)/4$ and $\alpha_{3,0}=\pi/4$, and (b)
errors in the parameters $\phi_2$ and $\phi_3$, i.e. deviations from their respective optimal values $\phi_{2,0} 
= 0$ and $\phi_{3,0}=\pi/2$.}
\end{figure}

\begin{figure}[t!]
\includegraphics[width=0.95\linewidth]{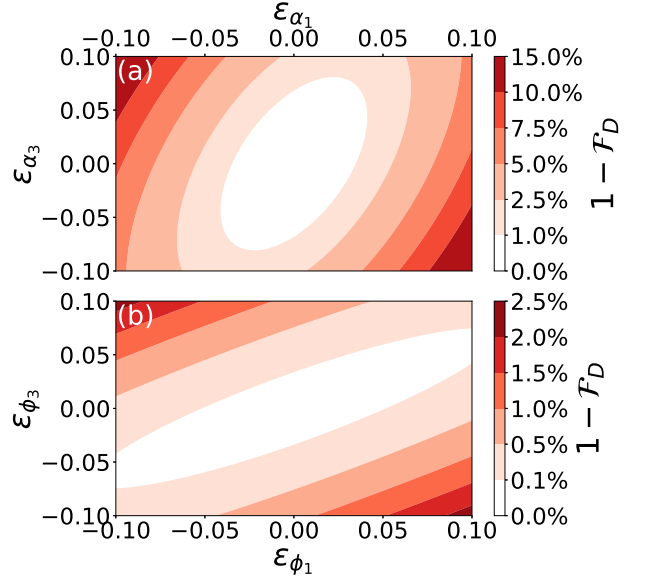}
\caption{\label{fig:InfidelityDensPlotTwoParam3}(Color online)Dependence of the deviation $1-\mathcal{F}_\text{D}$ 
of the Dicke-state fidelity from unity on (a) errors in the parameters $\alpha_1$ and $\alpha_3$, i.e. 
deviations from their respective optimal values $\alpha_{1,0}=3\pi/4$ and $\alpha_{3,0}=\pi/4$, and (b)
errors in the parameters $\phi_1$ and $\phi_3$, i.e. deviations from their optimal values $\phi_{1,0}=
\phi_{3,0}=\pi/2$.}
\end{figure}

\begin{figure}[t!]
\includegraphics[width=0.95\linewidth]{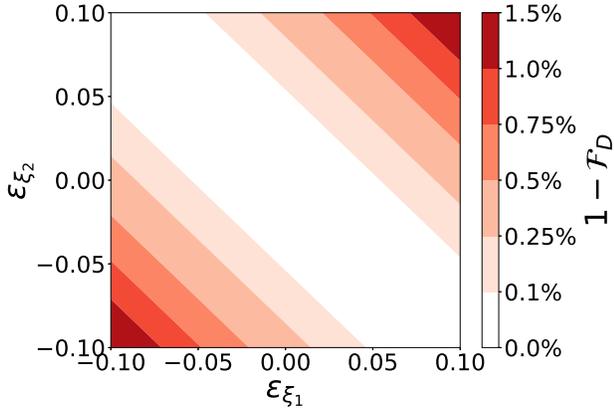}
\caption{\label{fig:InfidelityDensPlotTwoParam4}(Color online)Dependence of the deviation $1-\mathcal{F}_\text{D}$ 
of the Dicke-state fidelity from unity on errors in the Ising-pulse durations $\xi_1$ and $\xi_2$, i.e. 
deviations from their optimal values $\xi_{1,0}=\xi_{2,0}=[\pi-\arccos(1/3)]/4$.}
\end{figure}

Apart from the impacts of deviations $\varepsilon_p$ in the individual pulse-sequence parameters on the 
resulting Dicke-state fidelity, as illustrated by Fig.~\ref{fig:InfidelityVarParams}, it is pertinent 
to also quantitatively analyze the effect of simultaneous errors in more than one pulse-sequence parameter. 
To achieve that, we numerically evaluated the fidelity $\mathcal{F}_\text{D}$ in the presence of simultaneous 
errors in two different parameters, based on its defining expression [cf. Eq.~\eqref{DickeFidelity}]. The 
results obtained in this manner for the infidelity $1-\mathcal{F}_\text{D}$ are presented in the form of 
two-dimensional density plots in Figs.~\ref{fig:InfidelityDensPlotTwoParam1} - \ref{fig:InfidelityDensPlotTwoParam4}.
The near-elliptical shape of different regions in these plots originates from the fact that the dependence 
of $1-\mathcal{F}_\text{D}$ on the errors $\varepsilon_p$ is to leading order quadratic because these errors
represent deviations from the optimal parameter values.

For instance, Fig.~\ref{fig:InfidelityDensPlotTwoParam1}(a) shows the deviation $1-\mathcal{F}_\text{D}$ of the 
fidelity from unity in the presence of simultaneous errors in the parameters $\alpha_1$ and $\alpha_2$. Apart 
from corroborating the aforementioned conclusion that the target-state fidelity is much more sensitive to the 
deviation in the first rotation angle ($\alpha_1$) than in the second one ($\alpha_2$), another interesting quantitative 
insight can be gleaned from this density plot. Namely, it can be inferred from this plot that, as long as the relative 
errors in these two parameters are below $5\%$, the infidelity $1-\mathcal{F}_\text{D}$ does not exceed $2\%$, 
i.e. the Dicke-state fidelity is at least $0.98$. 

Figure~\ref{fig:InfidelityDensPlotTwoParam1}(b) illustrates the dependence of the infidelity on the deviations
from the optimal values of the parameters $\phi_1$ and $\phi_2$. What is immediately noticeable from this density 
plot is that the relevant infidelities are, roughly speaking, an order of magnitude lower than in the previously 
considered case of $\alpha_1$ and $\alpha_2$ [cf. Fig.~\ref{fig:InfidelityDensPlotTwoParam1}(a)]. For instance,
with the exception of very large deviations of $\phi_1$ and $\phi_2$ in the same direction (i.e. $\varepsilon_{\phi_1}$ 
and $\varepsilon_{\phi_2}$ have the same sign), the infidelity does not exceed $0.8\:\%$; in other words, the 
Dicke-state fidelity is above $0.99$ in almost the entire range of errors discussed. An important qualitative 
observation is that -- in contrast to the dependence of the infidelity on errors in the last two parameters, which
is completely symmetric with respect to the change of their sign -- the dependence of $1-\mathcal{F}_\text{D}$ 
on $\varepsilon_{\phi_{1}}$ and $\varepsilon_{\phi_{2}}$ is highly asymmetric with respect to such a change of sign.

The dependence of the infidelity on the errors in the parameters $\alpha_2$ and $\alpha_3$, which is illustrated 
in Fig.~\ref{fig:InfidelityDensPlotTwoParam2}(a), is qualitatively similar to that of $\alpha_1$ and $\alpha_2$
[cf. Fig.~\ref{fig:InfidelityDensPlotTwoParam1}(a)]. The main quantitative difference is that the elliptical regions
in Fig.~\ref{fig:InfidelityDensPlotTwoParam2}(a) are somewhat less elongated along the $\varepsilon_{\alpha_{2}}$ 
axis than their counterparts in Fig.~\ref{fig:InfidelityDensPlotTwoParam1}(a). In other words, the aspect ratio 
of the relevant elliptical curves is smaller than in Fig.~\ref{fig:InfidelityDensPlotTwoParam1}(a), which is yet 
another reflection of the dominant role of errors in the parameter $\alpha_1$ compared to errors in other pulse-sequence 
parameters. This trend is also evident in Fig.~\ref{fig:InfidelityDensPlotTwoParam3}(a), where the effect of simultaneous 
errors in the parameters $\alpha_1$ and $\alpha_3$ is illustrated. Another effect noticeable in the case of the latter
pair of parameters is the tilted orientation of the elliptical regions. 

Figure~\ref{fig:InfidelityDensPlotTwoParam2}(b) illustrates the dependence of the infidelity $1-\mathcal{F}_\text{D}$ 
on the simultaneous errors in the parameters $\phi_2$ and $\phi_3$, while Fig.~\ref{fig:InfidelityDensPlotTwoParam3}(b) 
depicts similar dependence on the errors in $\phi_1$ and $\phi_3$. Compared to Fig.~\ref{fig:InfidelityDensPlotTwoParam1}(b)
the relevant infidelities are somewhat larger in the latter two cases. However, the target-state fidelities are still
very close to unity -- more precisely, higher than $0.98$ and $0.975$, respectively -- even for the largest considered 
values of errors in those parameters. This speaks in favor of the robustness of the proposed state-preparation protocol.

By contrast to the rotation angles $\alpha_1,\alpha_2,\alpha_3$, the simultaneous errors in the parameters $\phi_1,\phi_2,\phi_3$ 
exhibit elliptical regions that are mainly elongated in the diagonal- and anti-diagonal directions. To provide a plausible explanation 
for this observation, we note that the entire system is axially symmetric, i.e. invariant under rotations around the $z$-axis. 
More precisely, the $x$- and $y$ axes can be chosen arbitrarily and a rotation around the $z$ axis merely leads to a physically 
irrelevant complex phase of the Dicke state. In particular, the Dicke-state fidelity ought to be invariant under uniform translations, 
i.e. transformations $\phi_1,\phi_2,\phi_3 \mapsto \phi_1+\epsilon,\phi_2+\epsilon,\phi_3+\epsilon$ for an arbitrary $\epsilon\in\mathbb{R}$. 
Needless to say, this invariance is not borne out by the obtained results as they correspond to the case of two simultaneous errors instead 
of three. However, this invariance makes the predominant orientation of the elliptical regions in the diagonal and anti-diagonal
directions plausible. For instance, the aforementioned symmetry implies that an error in the diagonal direction, that is,
the same errors in two angles (e.g., $\phi_{0,1},\phi_{0,2}+\epsilon,\phi_{0,3}+\epsilon$), coincides with an error in the 
third angle (e.g. $\phi_{0,1}-\epsilon,\phi_{0,2},\phi_{0,3}$). This last observation allows one to compare 
simultaneous errors in two angles $\phi_i$ and $\phi_j$ in the diagonal direction to an error in the remaining angle $\phi_k$ 
($i,j,k = 1,2,3$). In particular, Figure~\ref{fig:InfidelityVarParams} shows that the fidelity is most sensitive to deviations 
in $\phi_3$ (up to around $0.9\:\%$). This matches with the elliptical regions pertaining to deviations in $\phi_1$ and $\phi_2$ 
in Fig.~\ref{fig:InfidelityDensPlotTwoParam1}(b), which are oriented in the anti-diagonal direction; thus, the fastest increase 
of the infidelity is the one in the diagonal direction (also up to about $0.9\:\%$). Conversely, the elliptical regions
corresponding to deviations in $(\phi_1,\phi_3)$ and $(\phi_2,\phi_3)$ in Figs.~\ref{fig:InfidelityDensPlotTwoParam2}(b) and 
\ref{fig:InfidelityDensPlotTwoParam3}(b) are tilted in the diagonal direction, which corresponds to deviations in $\phi_2$ 
and $\phi_1$, respectively. Consistently, the infidelity increases slower in the diagonal direction than along the $\phi_3$ axis.

As can be inferred from Fig.~\ref{fig:InfidelityDensPlotTwoParam4}, the fidelity appears to remain equal to unity for anti-diagonal 
errors in $\xi_1$ and $\xi_2$, i.e. for $\xi_1=\xi_{1,0}+\epsilon$ and $\xi_2=\xi_{2,0}-\epsilon$ for $\epsilon\in\mathbb{R}$. 
This somewhat surprising observation should, however, be taken with caution because the fidelity does, in fact, deviate from unity 
for anti-diagonal errors, even if this deviation is hardly noticeable. Notwithstanding, we note that this observation is physically 
reasonable given that the total duration $T$ of the pulse sequence remains invariant under anti-diagonal errors in $\xi_1$ and 
$\xi_2$, this being a consequence of the fact that the total duration is proportional to $\xi_1+\xi_2$.

\section{Generalization: Dicke-state preparation for $N\ge 4$}   \label{GeneralizeNgt4}
For the sake of completeness, it is worthwhile explaining how the proposed scheme for the generation of the state 
$\ket{D^{3}_{2}}$ can be generalized to other Dicke states, in particular in systems with $N\ge 4$ qubits. To this 
end, we start this section with some general, permutational-symmetry related considerations (Sec.~\ref{GeneralConsider}). 
We then present the resulting pulse sequence for realizing the two-excitation Dicke state in a four-qubit system, 
i.e. the state $\ket{D^{4}_{2}}$ (Sec.~\ref{PulseSeqD42}).

\subsection{General permutational-symmetry-based considerations}   \label{GeneralConsider} 
To begin with, it is worthwhile noting that in an $N$-qubit system the subspace of permutationally-invariant 
states corresponds to the total spin of $N/2$; the simplest example is furnished by the $N=2$ case, i.e. that 
of two spin-$1/2$ particles, where the permutationally-invariant subspace corresponds to the total spin of $1$
and is three-dimensional (the spin-triplet subspace). Accordingly, the dimension of this permutationally-invariant 
subspace is equal to $N+1$~\cite{Stockton+:03}, hence increasing linearly with the number of qubits. This very 
fact -- that the dimension of the relevant permutationally-invariant subspace grows only linearly, and that of 
the total Hilbert space exponentially (i.e. as $2^N$), with the number of qubits -- speaks in favor of exploiting 
the permutational symmetry in the context of quantum-state engineering in systems of this type. 

One commonly used complete orthonormal basis of the permutationally-invariant subspace of the $N$-qubit Hilbert 
space is given by the Dicke states $\ket{D^{N}_{a}}$ ($a=0,\ldots,N$) [cf. Eq.~\eqref{DickeStateDef}]. In other 
words, any permutationally-invariant state of $N$ qubits can be expressed as a linear combination of the $N+1$ 
states of this symmetry-adapted basis; as a reminder, the basis in Eq.~\eqref{SSbasis} is the special $N=3$ case 
of this general Dicke-state basis. 

For an arbitrary qubit number $N$, it is straightforward to obtain the matrices representing 
the time-evolution operators $U_{ZZ}(\xi_j)$ and $U_C(\boldsymbol\alpha_j)$ in the chosen 
symmetry-adapted (Dicke-state) basis. To derive the matrix form of $U_{ZZ}(\xi_j)$, we first 
note that $H_{ZZ}$ is diagonal for each value of $N$. The corresponding matrix elements
of $H_{ZZ}$ are given by
\begin{equation}
\bra{D^N_a}H_{ZZ}\ket{D^N_a} = 2\left(a-\frac{N}{2}\right)^2-\frac{N}{2} \:,
\end{equation}
where $a=0,\ldots,N$. Hence, a simple matrix exponentiation yields
\begin{equation}
\bra{D^N_a}U_{ZZ}(\xi_j)\ket{D^N_a} = 
e^{-i\xi_j\left[2\left(a-{N}/{2}\right)^2-{N}/{2}\right]}.
\end{equation}
On the other hand, the time-evolution operator $U_C(\boldsymbol\alpha_j)$ corresponding 
to the $j$-th control pulse ($j=1,2,\ldots$) is given by
\begin{equation}
U_C(\boldsymbol\alpha_j) = e^{-i\alpha_j\mathcal{S}^{(j)}_1} \:,
\end{equation}
where the operator $\mathcal{S}^{(j)}_1$ is defined as [cf. Eq.~\eqref{ControlOperators}]
\begin{equation}
\mathcal{S}^{(j)}_1 = \cos\phi_j\mathcal{X} +
\sin\phi_j \mathcal{Y}  \:.
\end{equation}
It is straightforward to demonstrate that the matrix elements of $\mathcal{S}^{(j)}_1$ 
in the Dicke-state basis are given by
\begin{equation}
\begin{split}
\bra{D^N_a}\mathcal{S}^{(j)}_1\ket{D^N_b} =&
(\delta_{b,a+1}e^{i\phi_j}+\delta_{b,a-1}e^{-i\phi_j})
\\&\times
\sqrt{\frac{b+a+1}{2}\left(N-\frac{b+a-1}{2}\right)}  \:,
\end{split}
\end{equation}
for $a,b=0,\ldots,N$. Because $\mathcal{S}^{(j)}_1$ is not diagonal in the chosen basis, one 
can compute $U_C(\boldsymbol\alpha_j)$ using matrix exponentiation, which even in the most general
case can efficiently be carried out numerically.

Further insight into the problem of generating Dicke states in the $N$-qubit case can be gleaned 
by noting that the operators $H_{ZZ}$ and $\mathcal{X}$ commute with the $x$-parity operator 
$X_p\equiv X\otimes X\otimes\ldots\otimes X$. Consequently, these two operators can be block-diagonalized
in the two eigensubspaces of $X_p$ that correspond to the eigenvalues $\pm 1$; the bases of these
two eigensubspaces of $X_p$ are given by
\begin{equation} \label{XparityEigenv}
|v^{\pm}_k\rangle = c_k (\ket{D^{N}_{k-1}}\pm \ket{D^{N}_{N-k+1}}) \:,
\end{equation}
where $c_k$ is the normalization constant and $k=1,\ldots,\lfloor N/2 \rfloor+1$ (where, for a real 
number $x$, $\lfloor x \rfloor$ is the largest integer not larger than $x$). Note that for even values
of $N$ and the corresponding values of $k$ such that the two subscripts of Dicke states in Eq.~\eqref{XparityEigenv} 
are the same [i.e. for $N=2(k-1)$] the basis vector $|v^{-}_k\rangle$ vanishes; therefore, in that case
the $+1$ and $-1$ eigensubspaces of the ($N+1$)-dimensional permutationally-invariant subspace of the 
total $N$-qubit Hilbert space have dimensions equal to $1+N/2$ and $N/2$, respectively.

Similarly, the operators $H_{ZZ}$ and $\mathcal{Y}$ commute with the $y$-parity operator $Y_p\equiv Y
\otimes Y \otimes \ldots \otimes Y$ and -- as a result -- can both be block-diagonalized in the $\pm 1$ 
eigensubspaces of $Y_p$. The respective bases of these two eigensubspaces of $Y_p$ are given by
\begin{equation} \label{YparityEigenv}
|w^{\pm}_k\rangle = c_k [\ket{D^{N}_{k-1}}\pm i^{N}(-1)^{k}\ket{D^{N}_{N-k+1}}] \:,
\end{equation}
where $c_k$ and $k$ have the same meaning as in Eq.~\eqref{XparityEigenv} above.

\subsection{Pulse sequence for realizing the state $\ket{D^{4}_{2}}$}   \label{PulseSeqD42}
Having discussed permutational-symmetry-related aspects of the Dicke-state preparation in an $N$-qubit 
system in Sec.~\ref{GeneralConsider}, we now proceed to illustrate this general approach on the example 
of the two-excitation Dicke state in a four-qubit system, i.e. the state $\ket{D^{4}_{2}}$.
As we demonstrate in what follows, similar to the case of $\ket{D^{3}_{2}}$ the state $\ket{D^{4}_{2}}$ 
can also be realized starting from the state $|0000\rangle$ through a pulse sequence that consists 
of three global-control pulses and two Ising-interaction pulses, i.e. a pulse sequence of the 
type illustrated in Fig.~\ref{fig:PulseSequence}.

In the four-qubit case ($N=4$), the five relevant permutationally-invariant basis states are 
$\ket{D^{4}_{0}}\equiv |0000\rangle$, $\ket{D^{4}_{1}}$, $\ket{D^{4}_{2}}$, $\ket{D^{4}_{3}}$, 
and $\ket{D^{4}_{4}}\equiv |1111\rangle$ [for the explicit form of these states, see Eq.~\eqref{FourQubitBasis}].

In order to determine the optimal values of the pulse-sequence parameters, we optimize the 
target-state fidelity, which is given by an analog of Eq.~\eqref{DickeFidelity}. Namely, 
this state fidelity in the four-qubit case is given by
\begin{equation} \label{DickeFidelitySecond}
\mathcal{F}_{\textrm{D}}=\left|\bra{D^4_2}\,\mathcal{U}_{\textrm{PS}}(\xi_1,\xi_2,
\boldsymbol{\alpha}_1,\boldsymbol{\alpha}_2,\boldsymbol{\alpha}_3)\ket{0000}\right| \:.
\end{equation}
The optimal values of the eight pulse-sequence parameters, obtained numerically using the 
method previously employed in the three-qubit case (cf. Sec.~\ref{PulseSequence}), are 
\begin{eqnarray} \label{OptValuesParamsSecond}
\alpha_{1,0} &=& \pi/4  \:, \quad \alpha_{2,0} = - 1.162 \:, \quad \alpha_{3,0} = - 2.484 \:,\nonumber\\
\phi_{1,0} &=& \pi/2  \:, \quad \phi_{2,0} = \phi_{3,0} = 0\:, \\
\xi_{1,0} &=& 0.285 \:, \quad \xi_{2,0} = 0.702 \:. \nonumber 
\end{eqnarray}
Based on the obtained optimal parameter values one can conclude that the total duration 
$T=(\xi_{1,0}+\xi_{2,0})\:J^{-1}$ of the pulse sequence for realizing the state $\ket{D^4_2}$
is (upon reinstating $\hbar$) given by $T\approx 0.99\:\hbar/J$, which is just slightly 
longer than the time needed to realize the state $\ket{D^3_2}$ [cf. Eq.~\eqref{PSduration}].
Another, qualitative difference from the pulse sequence used for the preparation of $\ket{D^3_2}$
[cf. Sec.~\ref{PulseSequence}] is that in the four-qubit case the two Ising-interaction pulses 
do not have equal durations (i.e. $\xi_{1,0}\neq \xi_{2,0}$).

The obtained optimal values of the parameters $\alpha_{1}$ and $\phi_{j}\:(j=1,2,3)$ can be made
plausible through the following algebraic analysis. To begin with, we recall that the operators 
$H_{ZZ}$ and $\mathcal{X}$ commute with the parity operator $X_p$, implying that the eigensubspaces 
of $X_p$ are invariant under the action of $H_{ZZ}$ and $\mathcal{X}$ (cf. Sec.~\ref{GeneralConsider}). 
The eigensubspaces of $X_p$ corresponding to the eigenvalues $+1$ and $-1$ are spanned by the 
vectors $\{\ket{v^+_1},\ket{v^+_2},\ket{v^+_3}\}$ and $\{\ket{v^-_1},\ket{v^-_2}\}$, respectively, 
where [cf. Eq.~\eqref{XparityEigenv}]
\begin{equation}
\ket{v^\pm_1} = \frac{1}{\sqrt{2}}
\begin{pmatrix} 1\\0\\0\\0\\\pm 1 \end{pmatrix}
\,,\:\:
\ket{v^\pm_2} = \frac{1}{\sqrt{2}}
\begin{pmatrix} 0\\1\\0\\\pm 1\\0 \end{pmatrix}
\,,\:\:
\ket{v^+_3} = 
\begin{pmatrix} 0\\0\\1\\0\\0 \end{pmatrix} \:.
\end{equation}
In keeping with the general conclusion stated in Sec.~\ref{GeneralConsider} above, the 
respective dimensions of the $+1$ and $-1$ eigensubspaces of $X_p$ are $3$ and $2$. 

It is straightforward to first verify that $\ket{D^4_2}=\ket{v^+_3}$, i.e. the target state 
of our state-preparation scheme belongs to the subspace spanned by the vectors $\{\ket{v^+_1},\ket{v^+_2},
\ket{v^+_3}\}$. Therefore, it suffices to first convert the initial state $|0000\rangle\equiv\ket{D^4_0}$  
into a state that belongs to this last subspace, thereby effectively reducing the state-preparation 
problem at hand to the same subspace. It turns out that the desired initial state conversion 
requires a control pulse $U_C(\boldsymbol\alpha_{1,0})\equiv U_C(\alpha_{1,0},\phi_{1,0})$ with 
$\alpha_{1,0}=\pi/4$ and $\phi_{1,0}=\pi/2$, which corresponds to a rotation through an angle 
of $\pi/2$ around the $y$ axis. By first expressing $U_C(\alpha_{1,0}=\pi/4,\phi_{1,0}=\pi/2)$ 
in the symmetry-adapted basis, by means of Eq.~\eqref{eq:opS_symm-sec}, we obtain
\begin{equation}
U_C(\boldsymbol\alpha_{1,0}) \ket{D^4_0} = \frac{1}{4}
\begin{pmatrix} 1\\2\\\sqrt{6}\\2\\1 \end{pmatrix}
= \frac{\ket{v^+_1}+2\ket{v^+_2}+\sqrt{3}\ket{v^+_3}}{2\sqrt{2}} \:.
\end{equation}
This allows one to reduce the subsequent control pulses $U_C(\boldsymbol\alpha_{2,0})$ and 
$U_C(\boldsymbol\alpha_{3,0})$ to the subspace spanned by $\{\ket{v^+_1},\ket{v^+_2},\ket{v^+_3}\}$, 
which reveals that these control pulses correspond to global rotations around the $x$ axis, 
i.e. $\phi_{2,0}=\phi_{3,0}=0$. The remaining optimization problem -- i.e. the maximization 
of the state fidelity in Eq.~\eqref{DickeFidelitySecond} with respect to the remaining four 
pulse-sequence parameters, leading to the results given by Eq.~\eqref{OptValuesParamsSecond} -- can 
also be reduced by projecting the operators $H_{ZZ}$ and $\mathcal{X}$ into the same subspace:
\begin{equation}
H_{ZZ} \mapsto J \begin{pmatrix}
6&0&0 \\ 0&0&0 \\ 0&0&-2
\end{pmatrix}
\,,\quad
\mathcal{X} \mapsto 2\begin{pmatrix}
0&1&0 \\ 1&0&\sqrt{3} \\ 0&\sqrt{3}&0
\end{pmatrix}\:.
\end{equation}

\section{Summary and Conclusions} \label{SummConcl}
In summary, in this paper we addressed the problem of deterministically preparing the two-excitation Dicke state 
in a system that consists of three all-to-all Ising-coupled qubits acted upon by global control fields in the 
transverse directions. This system is state-to-state controllable for an arbitrary pair 
of initial and final states that are invariant with respect to permutations of qubits~\cite{Albertini+DAlessandro:18}. 
The permutational invariance of the Dicke state allowed us to carry out our analysis within the four-dimensional 
subspace of the three-qubit Hilbert space that consists of such (permutationally-invariant) states. 

We found a solution of the Dicke-state preparation problem that has the form of a five-stage NMR-type pulse 
sequence, which comprises three instantaneous control pulses -- equivalent to global qubit rotations -- and 
two Ising-interaction pulses. Through numerical optimization of the Dicke-state fidelity, we determined the 
optimal values of the eight parameters characterizing the envisioned pulse sequence. We then investigated the 
robustness of the proposed pulse sequence to systematic errors, i.e. deviations from the optimal values of 
those parameters. Importantly, we demonstrated that the Dicke-state fidelity remains very close to unity even 
for fairly large deviations from the optimal values of the relevant pulse-sequence parameters.

We also explained how our proposed scheme for the preparation of Dicke states can be generalized 
to systems with $N \ge 4$ qubits, describing -- as an example -- the preparation of the two-excitation Dicke state 
$|D^{4}_{2}\rangle$ in a four-qubit system. Likewise, the preparation of Dicke states could be addressed for qubit 
arrays with other relevant types of qubit-qubit interaction; for example, one could investigate this problem 
for an $XY$-type interaction, which is characteristic of superconducting qubits~\cite{StojanovicToffoli:12,
Stojanovic+:14,Stojanovic+Salom:19,Nauth+Stojanovic:23}, as well as for Heisenberg-type interactions that are of 
relevance for spin qubits~\cite{Heule+:10,Heule+EPJD:10,Stojanovic:19}. Experimental realization of the proposed 
state-preparation scheme in a physical platform with Ising-type coupling between qubits is keenly anticipated.

\begin{acknowledgments}
This research was supported by the Deutsche Forschungsgemeinschaft (DFG) -- SFB 1119 -- 236615297.
\end{acknowledgments}

\appendix
\section{Derivation of the relevant time-evolution operators} \label{DerivTimeEvol}
In the following, we sketch the derivation of the time-evolution operators corresponding to 
the control- and Ising-interaction pulses required for the preparation of the Dicke state $\ket{D^{3}_{2}}$
in a three-qubit ($N=3$) system (see Sec.~\ref{DerivNeq3} below), as well as
its counterpart $\ket{D^{4}_{2}}$ in a four-qubit ($N=4$) one (Sec.~\ref{DerivNeq4}).

\subsection{$N=3$ case} \label{DerivNeq3}
We start by representing the Ising-interaction Hamiltonian of a three-qubit ($N=3$) system 
[cf. Eq.~\eqref{threeQubitIsing}] in the symmetry-adapted basis of Eq.~\eqref{BasisColumnVecs}:
\begin{equation}
H_{ZZ} \mapsto J\:\begin{pmatrix}3 & 0 & 0 & 0 \\ 0 & -1 & 0 & 0\\ 0 & 0 & -1 & 0\\ 0 & 0 & 0 & 3 
\end{pmatrix} \:. \label{HzzMatrix}\\
\end{equation}
Given that $H_{ZZ}$ is already diagonal in the chosen basis, it is straightforward to write 
the explicit form of the time-evolution operators $U_{ZZ}(\xi_i)$ corresponding to both Ising 
interaction pulses $(i=1,2)$. Namely, these time-evolution operator are given by
\begin{equation} \label{TimeEvolutionIsing}
U_{ZZ}(\xi_i) = e^{-i\xi_i H_{ZZ}/J} \mapsto \begin{pmatrix}
e^{-3i\xi_i}&0&0&0\\0&e^{i\xi_i}&0&0\\0&0&e^{i\xi_i}&0\\0&0&0&e^{-3i\xi_i}
\end{pmatrix}\:,
\end{equation}
with $\xi_1\equiv JT_m$ being the dimensionless duration of the first interaction pulse
and $\xi_2\equiv J(T-T_m)$ that of the second one. 

We now turn to the derivation of the time-evolution operators $U_C(\boldsymbol{\alpha}_j)$ corresponding 
to the three instantaneous global control pulses ($j=1,2,3$) [cf. Eq.~\eqref{eq:h(t)_def}]. It is pertinent to 
first note that -- while the corresponding (time-dependent) control Hamiltonian [cf. Eq.~\eqref{threeQubitControl}] 
involves the mutually noncommuting Pauli operators $X_n$ and $Y_n$ of the $n$-th qubit ($n=1,2,3$) -- the $x$- 
and $y$ control fields have the same time-dependence. As a result, this control Hamiltonian commutes with itself 
at different times (i.e. $[H_C(t),H_C(t')]=0$). Accordingly, its corresponding time-evolution 
operator can be written in the simple form $\exp[-i\int_{t_i}^{t_f}H_C (t)dt]$ (with $t_i$ and $t_f$ being the 
initial- and final evolution times, respectively), rather than assuming the most-general form of a time-ordered 
exponential. More specifically yet, this time-evolution operator has the form of an exponential of a linear 
combination of the Pauli operators $X_n$ and $Y_n$ ($n=1,2,3$).  

By making use of the identity in Eq.~\eqref{PauliExpIdentity}, we arrive at the following 
expression for $U_C(\boldsymbol{\alpha}_j)\:(j=1,2,3)$:
\begin{equation}\label{eq:V_exponential}
U_C(\boldsymbol{\alpha}_j)=\prod_{n=1}^3\:[\cos\alpha_j\:\mathbbm{1}_{8}
-i\sin\alpha_j\:\mathcal{A}^{(j)}_n]  \:.
\end{equation}
The operators $\mathcal{A}^{(j)}_n$ ($n=1,2,3$) are here defined as
\begin{equation}\label{operatorAn}
\mathcal{A}^{(j)}_n=\frac{1}{\alpha_j}\:(\alpha_{j,x} X_n+\alpha_{j,y} Y_n) \:,
\end{equation}
where $\alpha_j\equiv \|\boldsymbol{\alpha}_j\|>0$ is the norm of the vector 
$\boldsymbol{\alpha}_j$, while $\alpha_{j,x}$ and $\alpha_{j,y}$ are its $x$- and $y$ 
components, respectively. The form of Eq.~\eqref{eq:V_exponential}, in conjunction with 
that of Eq.~\eqref{operatorAn}, makes it manifest that the three instantaneous global 
control pulses are equivalent to global qubit rotations.

In order to obtain an explicit form of the time-evolution operators $U_C(\boldsymbol{\alpha}_j)$, 
we proceed by performing the following two steps. Firstly, it should be noted that, using the polar 
coordinates in the $x$-$y$ plane the operator $\mathcal{A}^{(j)}_n$ -- acting on qubit $n$ -- can be 
recast in a simpler form. This is based on the identity 
\begin{equation}
\frac{1}{\alpha_j}\:(\alpha_{j,x} X+\alpha_{j,y} Y) = \begin{pmatrix}
0 & e^{-i\phi_j} \\ e^{i\phi_j} & 0
\end{pmatrix}\:
\end{equation}
for single-qubit Pauli operators, with $\phi_j$ being the polar angle corresponding 
to the vector $\boldsymbol{\alpha}_j$. Secondly, by carrying out the multiplication 
in Eq.~\eqref{eq:V_exponential} one obtains the expression 
\begin{eqnarray} \label{exprUc}
U_C(\boldsymbol{\alpha}_j) &=& \cos^3\alpha_j\:\mathbbm{1}_8
-i\sin\alpha_j\cos^2\alpha_j\:\mathcal{S}^{(j)}_1\nonumber\\
&-& \sin^2\alpha_j\cos\alpha_j\:\mathcal{S}^{(j)}_2+i\sin^3\alpha_j
\:\mathcal{S}^{(j)}_3 \:,
\end{eqnarray}
for $U_C(\boldsymbol{\alpha}_j)$, where the auxiliary operators $\mathcal{S}^{(j)}_1$, 
$\mathcal{S}^{(j)}_2$, and $\mathcal{S}^{(j)}_3$ ($j=1,2,3$) are obtained from 
$\mathcal{A}^{(j)}_n$ ($n=1,2,3$) using the following formulae:
\begin{eqnarray} \label{defmathcalS}
\mathcal{S}^{(j)}_1 &=& \sum_{n=1}^3 \mathcal{A}^{(j)}_n\:, \nonumber\\
\mathcal{S}^{(j)}_2 &=& \sum_{n<n'} \mathcal{A}^{(j)}_n\mathcal{A}^{(j)}_{n'}\:,\\\
\mathcal{S}^{(j)}_3 &=& \prod_{n=1}^3\mathcal{A}^{(j)}_n \:.\nonumber
\end{eqnarray}
When represented in the symmetry-adapted basis of Eq.~\eqref{BasisColumnVecs}, these operators 
are given by the $4\times 4$ matrices
\begin{eqnarray} \label{matricsmathcalS}
P_S \mathcal{S}^{(j)}_0 P_S^\dagger &=& \mathbbm{1}_{4}\:, \notag \\
P_S \mathcal{S}^{(j)}_1 P_S^{\dagger} &=& \begin{pmatrix}
0&\sqrt{3}\:e^{-i\phi_j}&0&0\\
\sqrt{3}\:e^{i\phi_j}&0&2e^{-i\phi_j}&0\\
0&2e^{i\phi_j}&0&\sqrt{3}\:e^{-i\phi_j}\\
0&0&\sqrt{3}\:e^{i\phi_j}&0
\end{pmatrix}\:,\notag\\
P_S \mathcal{S}^{(j)}_2 P_S^{\dagger} &=& \begin{pmatrix}
0&0&\sqrt{3}\:e^{-2i\phi_j}&0\\
0&2&0&\sqrt{3}\:e^{-2i\phi_j}\\
\sqrt{3}\:e^{2i\phi_j}&0&2&0\\
0&\sqrt{3}\:e^{2i\phi_j}&0&0
\end{pmatrix}\:,\notag\\
P_S \mathcal{S}^{(j)}_3 P_S^{\dagger} &=& \begin{pmatrix}
0&0&0&e^{-3i\phi_j}\\
0&0&e^{-i\phi_j}&0\\
0&e^{i\phi_j}&0&0\\
e^{3i\phi_j}&0&0&0
\end{pmatrix}\:,
\end{eqnarray}
where $P_S$ denotes the projector onto the symmetric sector [cf. Sec.~\ref{SymmSector}].
By inserting these last matrices into Eq.~\eqref{exprUc}, after elementary manipulations
one can obtain the matrices representing the time-evolution operators $U_C(\boldsymbol{\alpha}_j)$ 
in the chosen basis. The rather cumbersome final expressions will, however, not be provided 
explicitly here.

\subsection{$N=4$ case} \label{DerivNeq4}
Having presented a detailed derivation of the relevant time-evolution operators for a three-qubit
system (cf. Sec.~\ref{DerivNeq3}), in what follows we briefly sketch the derivation of their four-qubit 
($N=4$) counterparts. 

In the $N=4$ case the five (permutationally-invariant) Dicke basis states [cf. Eq.~\eqref{DickeStateDef}] 
are given by
\begin{eqnarray} 
\ket{D^{4}_{0}} &\equiv& |0000\rangle \:,\nonumber \\
\ket{D^{4}_{1}} &\equiv&  \frac{1}{2}\:
(|1000\rangle+|0100\rangle+|0010\rangle +|0001\rangle)\:, \nonumber \\
\ket{D^{4}_{2}} &\equiv& \frac{1}{\sqrt{6}}\:(|1100\rangle+|1010\rangle+|1001\rangle \nonumber \\
&+& |0110\rangle+|0101\rangle+|0011\rangle ) \:, \label{FourQubitBasis} \\
\ket{D^{4}_{3}} &\equiv&  \frac{1}{2}\:(|1110\rangle+|1101\rangle+|1011\rangle 
+|0111\rangle) \:, \nonumber \\
\ket{D^{4}_{4}} &\equiv&  |1111\rangle\:. \nonumber 
\end{eqnarray}
By first performing a mapping of these basis states onto column vectors, by analogy to Eq.~\eqref{BasisColumnVecs},
the Ising-interaction Hamiltonian of a four-qubit system is represented in the last symmetry-adapted basis as
\begin{equation}
H_{ZZ}\mapsto J \begin{pmatrix}
6 & 0 & 0 & 0 & 0 \\
0 & 0 & 0 & 0 & 0 \\
0 & 0 & -2 & 0 & 0 \\
0 & 0 & 0 & 0 & 0 \\
0 & 0 & 0 & 0 & 6
\end{pmatrix} \:.
\end{equation}
The form of the last equation directly leads us to conclude that 
the corresponding time-evolution operators $U_{ZZ}(\xi_i)$ ($i=1,2$) 
are given by
\begin{equation}
U_{ZZ}(\xi_i)
= e^{-i\xi_i H_{ZZ}/J}\mapsto \begin{pmatrix}
e^{-6i\xi_i} & 0 & 0 & 0 & 0 \\
0 & 1 & 0 & 0 & 0 \\
0 & 0 & e^{2i\xi_i} & 0 & 0 \\
0 & 0 & 0 & 1 & 0 \\
0 & 0 & 0 & 0 & e^{-6i\xi_i}
\end{pmatrix}\:.
\end{equation}
To determine the form of the time-evolution operators $U_C(\boldsymbol\alpha_j)$ of instantaneous global 
control pulses ($j=1,2,3$), we utilize the well-known identity in Eq.~\eqref{PauliExpIdentity}. 
In this way, we obtain the expression 
\begin{equation}
U_C(\boldsymbol\alpha_j) = \prod_{n=1}^4 [\cos\alpha_j \,\mathbbm{1}_{16}
- i\sin\alpha_j \mathcal{A}_n^{(j)}] \:,
\end{equation}
where the operators $\mathcal{A}_n^{(j)}$ are defined in Eq.~\eqref{operatorAn}. 
By making use of the binomial theorem, from the last equation we further obtain 
\begin{equation} \label{exprUcNeq4}
U_C(\boldsymbol\alpha_j) = \sum_{m=0}^4 (\cos\alpha_j)^{4-m}(-i\sin\alpha_j)^m
\mathcal{S}^{(j)}_m \:,
\end{equation}
where $\mathcal{S}^{(j)}_0 = \mathbbm{1}_{16}$ and the operators $\mathcal{S}^{(j)}_m$ 
($m=1,\ldots,4$) are constructed using the operators $\mathcal{A}_n^{(j)}$ according to 
\begin{equation}\label{eq:def_curly_S}
\mathcal{S}^{(j)}_m= \sum_{1\leq n_1<\ldots<n_m\leq 4}
\,\prod_{i=1}^m \mathcal{A}^{(j)}_{n_i} \:.
\end{equation}
It is worthwhile noting that the last equation generalizes Eq.~\eqref{defmathcalS} -- its 
three-qubit counterpart. 

The next step in the derivation of the time-evolution operators $U_C(\boldsymbol\alpha_j)$ 
corresponding to the instantaneous control pulses is to project the operators $\mathcal{S}^{(j)}_m$ 
onto the symmetry-adapted basis of Eq.~\eqref{FourQubitBasis}; we denote by $P_S$ the corresponding 
projector onto the five-dimensional permutationally-invariant subspace of the four-qubit Hilbert 
space. In this manner, we obtain the following $5\times 5$ matrices:
\begin{eqnarray}\label{eq:opS_symm-sec}
P_S \mathcal{S}^{(j)}_0 P_S^\dagger &=& \mathbbm{1}_{5}\:, \notag\\
P_S \mathcal{S}^{(j)}_1 P_S^\dagger &=&
\begin{pmatrix}
0&2e^{-i\phi_j}&0&0&0\\
2e^{i\phi_j}&0&\sqrt{6}\:e^{-i\phi_j}&0&0\\
0&\sqrt{6}\:e^{i\phi_j}&0&\sqrt{6}\:e^{-i\phi_j}&0\\
0&0&\sqrt{6}\:e^{i\phi_j}&0&2e^{-i\phi_j}\\
0&0&0&2e^{i\phi_j}&0
\end{pmatrix}\,, \notag\\
P_S \mathcal{S}^{(j)}_2 P_S^\dagger &=&
\begin{pmatrix}
0&0&\sqrt{6}\:e^{-2i\phi_j}&0&0\\
0&3&0&3e^{-2i\phi_j}&0\\
\sqrt{6}\:e^{2i\phi_j}&0&4&0&\sqrt{6}\:e^{-2i\phi_j}\\
0&3e^{2i\phi_j}&0&3&0\\
0&0&\sqrt{6}\:e^{2i\phi_j}&0&0
\end{pmatrix}\,, \notag \\
P_S \mathcal{S}^{(j)}_3 P_S^\dagger &=&
\begin{pmatrix}
0&0&0&2e^{-3i\phi_j}&0\\
0&0&\sqrt{6}\:e^{-i\phi_j}&0&2e^{-3i\phi_j}\\
0&\sqrt{6}\:e^{i\phi_j}&0&\sqrt{6}\:e^{-i\phi_j}&0\\
2e^{3i\phi_j}&0&\sqrt{6}\:e^{i\phi_j}&0&0\\
0&2e^{3i\phi_j}&0&0&0
\end{pmatrix}\,, \notag \\
P_S \mathcal{S}^{(j)}_4 P_S^\dagger &=&
\begin{pmatrix}
0&0&0&0&e^{-4i\phi_j}\\
0&0&0&e^{-2i\phi_j}&0\\
0&0&1&0&0\\
0&e^{2i\phi_j}&0&0&0\\
e^{4i\phi_j}&0&0&0&0
\end{pmatrix}\:. 
\end{eqnarray}

Finally, after inserting the obtained matrices into Eq.~\eqref{exprUcNeq4} and carrying out straightforward 
manipulations we can obtain the matrices representing the time-evolution operators $U_C(\boldsymbol{\alpha}_j)$ 
in the chosen symmetry-adapted basis.


\begin{thebibliography}{81}
\expandafter\ifx\csname natexlab\endcsname\relax\def\natexlab#1{#1}\fi
\expandafter\ifx\csname bibnamefont\endcsname\relax
  \def\bibnamefont#1{#1}\fi
\expandafter\ifx\csname bibfnamefont\endcsname\relax
  \def\bibfnamefont#1{#1}\fi
\expandafter\ifx\csname citenamefont\endcsname\relax
  \def\citenamefont#1{#1}\fi
\expandafter\ifx\csname url\endcsname\relax
  \def\url#1{\texttt{#1}}\fi
\expandafter\ifx\csname urlprefix\endcsname\relax\def\urlprefix{URL }\fi
\providecommand{\bibinfo}[2]{#2}
\providecommand{\eprint}[2][]{\url{#2}}

\bibitem[{Dog()}]{Dogra+:14}
\bibinfo{note}{S. Dogra, K. Dorai, and Arvind, Phys. Rev. A {\bf 91}, 022312
  (2014).}

\bibitem[{Das()}]{Das+:15}
\bibinfo{note}{D. Das, S. Dogra, K. Dorai, and Arvind, Phys. Rev. A {\bf 92},
  022307 (2015).}

\bibitem[{Li+()}]{Li+Song:15}
\bibinfo{note}{C. Li and Z. Song, Phys. Rev. A {\bf 91}, 062104 (2015).}

\bibitem[{Kan({\natexlab{a}})}]{Kang+:16}
\bibinfo{note}{Y.-H. Kang, Y.-H. Chen, Z.-C. Shi, J. Song, and Y. Xia, Phys.
  Rev. A {\bf 94}, 052311 (2016).}

\bibitem[{Kan({\natexlab{b}})}]{Kang+SciRep:16}
\bibinfo{note}{Y.-H. Kang, Y.-H. Chen, Q.-C. Wu, B.-H. Huang, J. Song, and Y.
  Xia, Sci. Rep. {\bf 6}, 36737 (2016).}

\bibitem[{Sto({\natexlab{a}})}]{StojanovicPRL:20}
\bibinfo{note}{V. M. Stojanovi\'c, Phys. Rev. Lett. {\bf 124}, 190504 (2020).}

\bibitem[{Pen()}]{Peng+:21}
\bibinfo{note}{J. Peng, J. Zheng, J. Yu, P. Tang, G. A. Barrios, J. Zhong, E.
  Solano, F. Albarr\'{a}n-Arriagada, and L. Lamata, Phys. Rev. Lett. {\bf 127},
  043604 (2021).}

\bibitem[{Sto({\natexlab{b}})}]{StojanovicPRA:21}
\bibinfo{note}{V. M. Stojanovi\'c, Phys. Rev. A {\bf 103}, 022410 (2021).}

\bibitem[{Zhe({\natexlab{a}})}]{Zheng++:22}
\bibinfo{note}{J. Zheng, J. Peng, P. Tang, F. Li, and N. Tan, Phys. Rev. A {\bf
  105}, 062408 (2022).}

\bibitem[{Zha({\natexlab{a}})}]{Zhang+:23}
\bibinfo{note}{G. Q. Zhang, W. Feng, W. Xiong, Q. P. Su, and C. P. Yang, Phys.
  Rev. A {\bf 107}, 012410 (2023).}

\bibitem[{Zha({\natexlab{b}})}]{Zhang++:23}
\bibinfo{note}{G. Q. Zhang, W. Feng, W. Xiong, D. Xu, Q. P. Su, and C. P. Yang,
  arXiv:2302.06204.}

\bibitem[{Son()}]{Song+:17}
\bibinfo{note}{C. Song, K. Xu, W. Liu, C.-P. Yang, S.-B. Zheng, H. Deng, Q.
  Xie, K. Huang, Q. Guo, L. Zhang, {\em et al.}, Phys. Rev. Lett. {\bf 119},
  180511 (2017).}

\bibitem[{Erh()}]{Erhard+:18}
\bibinfo{note}{M. Erhard, M. Malik, M. Krenn, and A. Zeilinger, Nat. Photon.
  {\bf 12}, 759 (2018).}

\bibitem[{Mac()}]{Macri+:18}
\bibinfo{note}{V. Macr{\`{i}}, F. Nori, and A. Frisk Kockum, Phys. Rev. A {\bf
  98}, 062327 (2018).}


\bibitem[{Zhe({\natexlab{b}})}]{Zheng+:19}
\bibinfo{note}{R.-H. Zheng, Y.-H. Kang, Z.-C. Shi, and Y. Xia, Ann. Phys.
  (Berlin) {\bf 531}, 1800447 (2019).}

\bibitem[{Nog()}]{Nogueira+:21}
\bibinfo{note}{J. Nogueira, P. A. Oliveira, F. M. Souza, and L. Sanz, Phys.
  Rev. A {\bf 103}, 032438 (2021).}

\bibitem[{Qia()}]{Qiao+:22}
\bibinfo{note}{Y.-F. Qiao, J.-Q. Chen, X.-L. Dong, B.-L. Wang, X.-L. Hei, C.-P.
  Shen, Y. Zhou, and P.-B. Li, Phys. Rev. A {\bf 105}, 032415 (2022).}

\bibitem[{Fen()}]{Feng+:22}
\bibinfo{note}{W. Feng, G. Q. Zhang, Q. P. Su, J. X. Zhang, and C. P. Yang,
  Phys. Rev. Applied {\bf 18}, 064036 (2022).}

\bibitem[{Pac()}]{Pachniak+Malinovskaya:21}
\bibinfo{note}{E. Pachniak and S. A. Malinovskaya, Sci. Rep. {\bf 11}, 12980 (2021).}

\bibitem[{Hau({\natexlab{a}})}]{Hauck+:21}
\bibinfo{note}{S. H. Hauck, G. Alber, and V. M. Stojanovi\'c, Phys. Rev. A {\bf
  104}, 053110 (2021).}

\bibitem[{Hau({\natexlab{b}})}]{Hauck+Stojanovic:22}
\bibinfo{note}{S. H. Hauck and V. M. Stojanovi\'c, Phys. Rev. Applied {\bf 18},
  014016 (2022).}

\bibitem[{Zhe({\natexlab{c}})}]{Zheng+:20}
\bibinfo{note}{R.-H. Zheng, Y.-H. Kang, D. Ran, Z.-C. Shi, and Y. Xia, Phys.
  Rev. A {\bf 101}, 012345 (2020).}

\bibitem[{Haa({\natexlab{a}})}]{Haase+:21}
\bibinfo{note}{T. Haase, G. Alber, and V. M. Stojanovi\'c, Phys. Rev. A {\bf
  103}, 032427 (2021).}

\bibitem[{Haa({\natexlab{b}})}]{Haase++:22}
\bibinfo{note}{T. Haase, G. Alber, and V. M. Stojanovi\'c, Phys. Rev. Research
  {\bf 4}, 033087 (2022).}

\bibitem[{Nau()}]{Nauth+Stojanovic:22}
\bibinfo{note}{J. K. Nauth and V. M. Stojanovi\'c, Phys. Rev. A {\bf 106},
  032605 (2022).}

\bibitem[{Sto({\natexlab{c}})}]{Stojanovic+Nauth:22}
\bibinfo{note}{V. M. Stojanovi\'c and J. K. Nauth, Phys. Rev. A {\bf 106},
  052613 (2022).}
  
\bibitem[{Sha({\natexlab{c}})}]{Shao+:23}
\bibinfo{note}{X. Q. Shao, F. Liu, X. W. Xue, W. L. Mu, and W. Li, 
arXiv:2303.13039.}  
  
\bibitem[{Dow()}]{Dowling+Milburn:03}
\bibinfo{note}{J. P. Dowling and G. J. Milburn, Phil. Trans. R. Soc. A {\bf
  361}, 1655 (2003).}

\bibitem[{Haa({\natexlab{c}})}]{Haas+:14}
\bibinfo{note}{F. Haas, J. Volz, R. Gehr, J. Reichel, and J. Esteve, Science
  {\bf 344}, 180 (2014).}

\bibitem[{Fri()}]{Friis+:18}
\bibinfo{note}{N. Friis {\em et al.}, Phys. Rev. X {\bf 8}, 021012 (2018).}

\bibitem[{Due()}]{Duer+:00}
\bibinfo{note}{W. D\"{u}r, G. Vidal, and J. I. Cirac, Phys. Rev. A {\bf 62},
  062314 (2000).}

\bibitem[{Gre()}]{Greenberger+Horne+Zeilinger:89}
\bibinfo{note}{D. M. Greenberger, M. A. Horne, and A. Zeilinger, in {\em Bell's
  Theorem, Quantum Theory, and Conceptions of the Universe} (Kluwer Academic,
  Dordrecht, 1989), pp. 73-76.}

\bibitem[{\citenamefont{Nielsen and Chuang}(2000)}]{NielsenChuangBook}
\bibinfo{author}{\bibfnamefont{M.~A.} \bibnamefont{Nielsen}} \bibnamefont{and}
  \bibinfo{author}{\bibfnamefont{I.~L.} \bibnamefont{Chuang}},
  \emph{\bibinfo{title}{{Q}uantum {C}omputation and {Q}uantum {I}nformation}}
  (\bibinfo{publisher}{Cambridge University Press},
  \bibinfo{address}{Cambridge}, \bibinfo{year}{2000}).

\bibitem[{Dic()}]{Dicke:54}
\bibinfo{note}{R. H. Dicke, Phys. Rev. {\bf 93}, 99 (1954).}

\bibitem[{Nev()}]{Neven+:18}
\bibinfo{note}{A. Neven, J. Martin, and T. Bastin, Phys. Rev. A {\bf 98},
  062335 (2018).}

\bibitem[{Lid()}]{Lidar+Whaley:03}
\bibinfo{note}{D. A. Lidar and K. B. Whaley, in {\em Irreversible Quantum
  Dynamics} (Springer, Berlin Heidelberg, 2003), pp. 83-120}.

\bibitem[{Pre()}]{Prevedel+:09}
\bibinfo{note}{R. Prevedel, G. Cronenberg, M. S. Tame, M. Paternostro, P.
  Walther, M. S. Kim, and A. Zeilinger, Phys. Rev. Lett. {\bf 103}, 020503
  (2009).}

\bibitem[{Tot()}]{Toth:12}
\bibinfo{note}{G. Toth, Phys. Rev. A ${\mathbf{85}}$, 022322 (2012).}

\bibitem[{Ozd()}]{Ozdemir+:07}
\bibinfo{note}{S. K. \"{O}zdemir, J. Shimamura, and N. Imoto, New J. Phys.
  ${\mathbf{9}}$, 43 (2007).}

\bibitem[{Chi()}]{Childs+:02}
\bibinfo{note}{A. M. Childs, E. Farhi, J. Goldstone, and S. Gutmann, Quantum
  Inf. Comput. ${\mathbf{2}}$, 181 (2002).}

\bibitem[{Hum()}]{Hume+:09}
\bibinfo{note}{D. B. Hume, C. W. Chou, T. Rosenband, and D. J. Wineland, Phys.
  Rev. A {\bf 80}, 052302 (2009).}

\bibitem[{\citenamefont{Ivanov et~al.}(2013)\citenamefont{Ivanov, Porras,
  Ivanov, and Schmidt-Kaler}}]{Ivanov+:13}
\bibinfo{author}{\bibfnamefont{P.~A.} \bibnamefont{Ivanov}},
  \bibinfo{author}{\bibfnamefont{D.}~\bibnamefont{Porras}},
  \bibinfo{author}{\bibfnamefont{S.~S.} \bibnamefont{Ivanov}},
  \bibnamefont{and}
  \bibinfo{author}{\bibfnamefont{F.}~\bibnamefont{Schmidt-Kaler}},
  \bibinfo{journal}{J. Phys. B: At. Mol. Opt. Phys.}
  \textbf{\bibinfo{volume}{46}}, \bibinfo{pages}{104003}
  (\bibinfo{year}{2013}).

\bibitem[{Lam()}]{Lamata+:13}
\bibinfo{note}{L. Lamata, C. E. L\'{o}pez, B. P. Lanyon, T. Bastin, J. C.
  Retamal, and E. Solano, Phys. Rev. A {\bf 87}, 032325 (2013).}

\bibitem[{Sto({\natexlab{d}})}]{Stockton+:04}
\bibinfo{note}{J. K. Stockton, R. van Handel, and H. Mabuchi, Phys. Rev. A {\bf
  70}, 022106 (2004).}

\bibitem[{Xia()}]{Xiao+:07}
\bibinfo{note}{Y.-F. Xiao, X.-B. Zou, and G.-C. Guo, Phys. Rev. A {\bf 75},
  012310 (2007).}

\bibitem[{Sha()}]{Shao+:10}
\bibinfo{note}{X.-Q. Shao, L. Chen, S. Zhang, Y.-F. Zhao, and K.-H. Yeon, EPL
  {\bf 90}, 50003 (2010).}

\bibitem[{Wie()}]{Wieczorek+:09}
\bibinfo{note}{W. Wieczorek, R. Krischek, N. Kiesel, P. Michelberger, G. Toth,
  and H. Weinfurter, Phys. Rev. Lett. {\bf 103}, 020504 (2009).}

\bibitem[{Wu+()}]{Wu+:17}
\bibinfo{note}{C. Wu, C. Guo, Y. Wang, G. Wang, X.-L. Feng, and J.-L. Chen,
  Phys. Rev. A {\bf 95}, 013845 (2017).}

\bibitem[{Wan()}]{Wang+Terhal:21}
\bibinfo{note}{Y. Wang and B. M. Terhal, Phys. Rev. A {\bf 104}, 032407
  (2021).}

\bibitem[{Sol()}]{Solgun+Srinivasan:22}
\bibinfo{note}{F. Solgun and S. Srinivasan, Phys. Rev. Applied {\bf 18}, 044025
  (2022).}

\bibitem[{Van()}]{Vandersypen+Chuang:05}
\bibinfo{note}{For a review, see L. M. K. Vandersypen and I. L. Chuang, Rev.
  Mod. Phys. {\bf 76}, 1037 (2005).}

\bibitem[{\citenamefont{Viola et~al.}(2001)\citenamefont{Viola, Knill, and
  Laflamme}}]{Viola+:01}
\bibinfo{author}{\bibfnamefont{L.}~\bibnamefont{Viola}},
  \bibinfo{author}{\bibfnamefont{E.}~\bibnamefont{Knill}}, \bibnamefont{and}
  \bibinfo{author}{\bibfnamefont{R.}~\bibnamefont{Laflamme}},
  \bibinfo{journal}{J. Phys. A: Math. Gen.} \textbf{\bibinfo{volume}{34}},
  \bibinfo{pages}{7067} (\bibinfo{year}{2001}).

\bibitem[{Zha({\natexlab{c}})}]{Zhang+:15}
\bibinfo{note}{J. Zhang, D. Burgarth, R. Laflamme, and D. Suter, Phys. Rev. A
  ${\mathbf{91}}$, 012330 (2015).}

\bibitem[{Shu()}]{Shulman+:12}
\bibinfo{note}{M. D. Shulman, O. E. Dial, S. P. Harvey, H. Bluhm, V. Umansky,
  and A. Yacoby, Science {\bf 336}, 202 (2012).}

\bibitem[{Lev()}]{Levy:02}
\bibinfo{note}{J. Levy, Phys. Rev. Lett. {\bf 89}, 147902 (2002).}

\bibitem[{Mor()}]{Morgado+Whitlock:21}
\bibinfo{note}{For an extensive review, see, e.g., M. Morgado and S. Whitlock,
  AVS Quantum Sci. {\bf 3}, 023501 (2021).}

\bibitem[{Shi()}]{ShiREVIEW:22}
\bibinfo{note}{For an up-to-date review, see X.-F. Shi, Quantum Sci. Technol.
  {\bf 7}, 023002 (2022).}
  
\bibitem[{LiP()}]{Li+:22}
\bibinfo{note}{X. X. Li, J. B. You, X. Q. Shao, and W. Li, 
Phys. Rev. A {\bf 105}, 032417 (2022).}

\bibitem[{Alb()}]{Albertini+DAlessandro:18}
\bibinfo{note}{F. Albertini and D. D'Alessandro, J. Math. Phys.
  ${\mathbf{59}}$, 052102 (2018).}

\bibitem[{Jon()}]{Jones:03}
\bibinfo{note}{J. A. Jones, Phys. Rev. A ${\mathbf{67}}$, 012317 (2003).}

\bibitem[{Hil()}]{Hill:07}
\bibinfo{note}{C. D. Hill, Phys. Rev. Lett. ${\mathbf{98}}$, 180501 (2007).}

\bibitem[{\citenamefont{Geller et~al.}(2009)\citenamefont{Geller, Pritchett,
  Galiautdinov, and Martinis}}]{Geller+:10}
\bibinfo{author}{\bibfnamefont{M.~R.} \bibnamefont{Geller}},
  \bibinfo{author}{\bibfnamefont{E.~J.} \bibnamefont{Pritchett}},
  \bibinfo{author}{\bibfnamefont{A.}~\bibnamefont{Galiautdinov}},
  \bibnamefont{and} \bibinfo{author}{\bibfnamefont{J.~M.}
  \bibnamefont{Martinis}}, \bibinfo{journal}{Phys. Rev. A}
  \textbf{\bibinfo{volume}{81}}, \bibinfo{pages}{012320}
  (\bibinfo{year}{2009}).

\bibitem[{\citenamefont{Ghosh and Geller}(2010)}]{Ghosh+Geller:10}
\bibinfo{author}{\bibfnamefont{J.}~\bibnamefont{Ghosh}} \bibnamefont{and}
  \bibinfo{author}{\bibfnamefont{M.~R.} \bibnamefont{Geller}},
  \bibinfo{journal}{Phys. Rev. A} \textbf{\bibinfo{volume}{81}},
  \bibinfo{pages}{052340} (\bibinfo{year}{2010}).

\bibitem[{Tan()}]{TanamotoQECC}
\bibinfo{note}{See, e.g, T. Tanamoto, Phys. Rev. A ${\mathbf{88}}$, 062334
  (2013).}

\bibitem[{\citenamefont{Tanamoto et~al.}(2012)\citenamefont{Tanamoto, Becker,
  Stojanovi\'c, and Bruder}}]{Tanamoto+:12}
\bibinfo{author}{\bibfnamefont{T.}~\bibnamefont{Tanamoto}},
  \bibinfo{author}{\bibfnamefont{D.}~\bibnamefont{Becker}},
  \bibinfo{author}{\bibfnamefont{V.~M.} \bibnamefont{Stojanovi\'c}},
  \bibnamefont{and} \bibinfo{author}{\bibfnamefont{C.}~\bibnamefont{Bruder}},
  \bibinfo{journal}{Phys. Rev. A} \textbf{\bibinfo{volume}{86}},
  \bibinfo{pages}{032327} (\bibinfo{year}{2012}).

\bibitem[{\citenamefont{Tanamoto et~al.}(2013)\citenamefont{Tanamoto,
  Stojanovi\'c, Bruder, and Becker}}]{Tanamoto+:13}
\bibinfo{author}{\bibfnamefont{T.}~\bibnamefont{Tanamoto}},
  \bibinfo{author}{\bibfnamefont{V.~M.} \bibnamefont{Stojanovi\'c}},
  \bibinfo{author}{\bibfnamefont{C.}~\bibnamefont{Bruder}}, \bibnamefont{and}
  \bibinfo{author}{\bibfnamefont{D.}~\bibnamefont{Becker}},
  \bibinfo{journal}{Phys. Rev. A} \textbf{\bibinfo{volume}{87}},
  \bibinfo{pages}{052305} (\bibinfo{year}{2013}).
  
\bibitem[{Ste()}]{Stefanatos+Paspalakis:20}
\bibinfo{note}{D. Stefanatos and E. Paspalakis, Phys. Rev. A
${\mathbf{102}}$, 052618 (2020).}  

\bibitem[{Rau()}]{Raussendorf+Briegel:01}
\bibinfo{note}{R. Raussendorf and H. J. Briegel, Phys. Rev. Lett.
  ${\mathbf{86}}$, 5188 (2001).}

\bibitem[{\citenamefont{D'Alessandro}(2008)}]{D'AlessandroBook}
\bibinfo{author}{\bibfnamefont{D.}~\bibnamefont{D'Alessandro}},
  \emph{\bibinfo{title}{Introduction to {Q}uantum {C}ontrol and {D}ynamics}}
  (\bibinfo{publisher}{Taylor \& Francis}, \bibinfo{address}{Boca Raton},
  \bibinfo{year}{2008}).

\bibitem[{\citenamefont{Wang et~al.}(2016)\citenamefont{Wang, Burgarth, and
  Schirmer}}]{Wang++:16}
\bibinfo{author}{\bibfnamefont{X.}~\bibnamefont{Wang}},
  \bibinfo{author}{\bibfnamefont{D.}~\bibnamefont{Burgarth}}, \bibnamefont{and}
  \bibinfo{author}{\bibfnamefont{S.~G.} \bibnamefont{Schirmer}},
  \bibinfo{journal}{Phys. Rev. A} \textbf{\bibinfo{volume}{94}},
  \bibinfo{pages}{052319} (\bibinfo{year}{2016}).

\bibitem[{\citenamefont{Zhang and Whaley}(2005)}]{Zhang+Whaley:05}
\bibinfo{author}{\bibfnamefont{J.}~\bibnamefont{Zhang}} \bibnamefont{and}
  \bibinfo{author}{\bibfnamefont{K.~B.} \bibnamefont{Whaley}},
  \bibinfo{journal}{Phys. Rev. A} \textbf{\bibinfo{volume}{71}},
  \bibinfo{pages}{052317} (\bibinfo{year}{2005}).

\bibitem[{Zan()}]{Zanardi:99}
\bibinfo{note}{P. Zanardi, Phys. Rev. A ${\mathbf{60}}$, R729 (1999).}

\bibitem[{Rib()}]{Ribeiro+Mosseri:11}
\bibinfo{note}{P. Ribeiro and M. Mosseri, Phys. Rev. Lett. ${\mathbf{106}}$,
  180502 (2011).}

\bibitem[{Bur()}]{Burchardt+:21}
\bibinfo{note}{A. Burchardt, J. Czartowski, and K. \.{Z}yczkowski, Phys. Rev. A
  ${\mathbf{104}}$, 022426 (2021).}

\bibitem[{Lyo()}]{Lyons+:22}
\bibinfo{note}{D. W. Lyons, J. R. Arnold, and A. F. Swogger, Phys. Rev. A
  ${\mathbf{105}}$, 032442 (2022).}
  
\bibitem[{Gall()}]{GallagherBOOK}
\bibinfo{note}{T. F. Gallagher, {\em Rydberg Atoms} (Cambridge University 
 Press, Cambridge, 1994).}   
  
\bibitem[{Sto()}]{Stockton+:03}
\bibinfo{note}{J. K. Stockton, J. M. Geremia, A. C. Doherty, and H. Mabuchi, 
Phys. Rev. A {\bf 67}, 022112 (2003).}  
  

\bibitem[{Kau()}]{Kaufman+Ni:21}
\bibinfo{note}{For a recent review on optical tweezer arrays, see A. M. 
Kaufman and K.-K. Ni, Nat. Phys. {\bf 17}, 1324 (2021).}

\bibitem[{min()}]{minimize_scipy}
\bibinfo{note}{For the reference material, see the following URL:
  https://docs.scipy.org/doc/scipy/reference/generated/\\
  scipy.optimize.minimize.html.}

\bibitem[{\citenamefont{Stojanovi{\'{c}}
  et~al.}(2012)\citenamefont{Stojanovi{\'{c}}, Fedorov, Wallraff, and
  Bruder}}]{StojanovicToffoli:12}
\bibinfo{author}{\bibfnamefont{V.~M.} \bibnamefont{Stojanovi{\'{c}}}},
  \bibinfo{author}{\bibfnamefont{A.}~\bibnamefont{Fedorov}},
  \bibinfo{author}{\bibfnamefont{A.}~\bibnamefont{Wallraff}}, \bibnamefont{and}
  \bibinfo{author}{\bibfnamefont{C.}~\bibnamefont{Bruder}},
  \bibinfo{journal}{Phys. Rev. B} \textbf{\bibinfo{volume}{85}},
  \bibinfo{pages}{054504} (\bibinfo{year}{2012}).

\bibitem[{\citenamefont{Stojanovi{\'{c}}
  et~al.}(2014)\citenamefont{Stojanovi{\'{c}}, Vanevi{\'{c}}, Demler, and
  Tian}}]{Stojanovic+:14}
\bibinfo{author}{\bibfnamefont{V.~M.} \bibnamefont{Stojanovi{\'{c}}}},
  \bibinfo{author}{\bibfnamefont{M.}~\bibnamefont{Vanevi{\'{c}}}},
  \bibinfo{author}{\bibfnamefont{E.}~\bibnamefont{Demler}}, \bibnamefont{and}
  \bibinfo{author}{\bibfnamefont{L.}~\bibnamefont{Tian}},
  \bibinfo{journal}{Phys. Rev. B} \textbf{\bibinfo{volume}{89}},
  \bibinfo{pages}{144508} (\bibinfo{year}{2014}).

\bibitem[{\citenamefont{Stojanovi{\'{c}} and
  Salom}(2019)}]{Stojanovic+Salom:19}
\bibinfo{author}{\bibfnamefont{V.~M.} \bibnamefont{Stojanovi{\'{c}}}}
  \bibnamefont{and} \bibinfo{author}{\bibfnamefont{I.}~\bibnamefont{Salom}},
  \bibinfo{journal}{Phys. Rev. B} \textbf{\bibinfo{volume}{99}},
  \bibinfo{pages}{134308} (\bibinfo{year}{2019}).

\bibitem[{NauthSto()}]{Nauth+Stojanovic:23}
\bibinfo{note}{J. K. Nauth and V. M. Stojanovi\'c, Phys. Rev. B 
${\mathbf{107}}$, 174306 (2023).}

\bibitem[{\citenamefont{Heule et~al.}(2010)\citenamefont{Heule, Bruder,
  Burgarth, and Stojanovi\'c}}]{Heule+:10}
\bibinfo{author}{\bibfnamefont{R.}~\bibnamefont{Heule}},
  \bibinfo{author}{\bibfnamefont{C.}~\bibnamefont{Bruder}},
  \bibinfo{author}{\bibfnamefont{D.}~\bibnamefont{Burgarth}}, \bibnamefont{and}
  \bibinfo{author}{\bibfnamefont{V.~M.} \bibnamefont{Stojanovi\'c}},
  \bibinfo{journal}{Phys. Rev. A} \textbf{\bibinfo{volume}{82}},
  \bibinfo{pages}{052333} (\bibinfo{year}{2010}).

\bibitem[{Heu()}]{Heule+EPJD:10}
\bibinfo{note}{R. Heule, C. Bruder, D. Burgarth, and V. M. Stojanovi\'c, Eur.
  Phys. J. D ${\mathbf{63}}$, 41 (2011).}

\bibitem[{Sto({\natexlab{e}})}]{Stojanovic:19}
\bibinfo{note}{See, e.g., V. M. Stojanovi\'{c}, Phys. Rev. A ${\mathbf{99}}$,
  012345 (2019).}

\end{thebibliography}
\end{document}